%% file: main.tex
\documentclass[aip,rse,amsmath,amssymb,preprint]{revtex4-1}

\usepackage[utf8]{inputenc}
\usepackage[T1]{fontenc}
\usepackage{mathptmx}
\usepackage{etoolbox}
\usepackage{graphicx}
\usepackage{dcolumn}
\usepackage{bm}
\usepackage[mathlines]{lineno}

\usepackage{xcolor}
\usepackage{newtxtext}
\usepackage{newtxmath}
\usepackage{makecell}

\usepackage{graphicx}
\usepackage[export]{adjustbox}

\usepackage{epstopdf} 

\usepackage{rotating}
\usepackage{adjustbox}
\usepackage{booktabs}
\usepackage{multirow} 
\usepackage{threeparttable} 
\usepackage{placeins}
\usepackage{tabularx}
\usepackage[]{natbib}
\usepackage{url}
\usepackage{hyperref}
\hypersetup{
    colorlinks = true,
    urlcolor   = blue,
    citecolor  = blue, 
    linkcolor = blue
}
\usepackage[capitalise]{cleveref} 

\usepackage[final,defaultcolor=red]{changes} 

\usepackage{enumitem}

\def\@email#1#2{%
 \endgroup
 \patchcmd{\titleblock@produce}
  {\frontmatter@RRAPformat}
  {\frontmatter@RRAPformat{\produce@RRAP{*#1\href{mailto:#2}{#2}}}\frontmatter@RRAPformat}
  {}{}
}%
\makeatother

\begin{document}

\preprint{}

\title[Performance characteristics and bluff-body modeling of high-blockage cross-flow turbine arrays...]{Performance characteristics and bluff-body modeling of high-blockage cross-flow turbine arrays with varying rotor geometry}
\author{Aidan Hunt}
    \email{ahunt94@uw.edu}
    \noaffiliation
    \affiliation{Department of Mechanical Engineering, University of Washington, \\ Seattle, WA 98195-2600, USA}%

\author{Gregory Talpey}%
    \noaffiliation
    \affiliation{Department of Mechanical Engineering, University of Washington, \\ Seattle, WA 98195-2600, USA}%
    
\author{Gemma Calandra}
    \noaffiliation
    \affiliation{Department of Mechanical Engineering, University of Washington, \\ Seattle, WA 98195-2600, USA}%

\author{Brian Polagye}
    \noaffiliation
    \affiliation{Department of Mechanical Engineering, University of Washington, \\ Seattle, WA 98195-2600, USA}%

\date{\today}

\begin{abstract}
    While confinement is understood to increase the power and thrust coefficients of cross-flow turbines, how the optimal rotor geometry changes with the blockage ratio---defined as the ratio between the array projected area and the channel cross-sectional area---has not been systematically explored.
    Here, the interplay between rotor geometry and the blockage ratio on turbine performance is investigated experimentally with an array of two identical cross-flow turbines at blockage ratios from $35\%$ to $55\%$.
    Three geometric parameters are varied---the number of blades, the chord-to-radius ratio, and the preset pitch angle---resulting in 180 unique combinations of rotor geometry and blockage ratio.
    While the optimal chord-to-radius ratio and preset pitch angle do not depend on the blockage ratio, the optimal blade count increases with the blockage ratio---an inversion of the relationship between efficiency and blade count typically observed at lower blockage.
    To explore the combined effects of rotor geometry, rotation rate, and the blockage ratio on array performance, we utilize two bluff-body models: dynamic solidity (which relates thrust to the rotor geometry and kinematics) and Maskell-inspired linear momentum theory (which describes the array-channel interaction as a function of the blockage ratio and thrust).
    By combining these models, we demonstrate that the array time-average thrust coefficient increases with dynamic solidity in a manner that is self-similar across blockage ratios.
    Overall, these results highlight key design principles for cross-flow turbines in confined flow and provide insights into the similarities between the dynamics of cross-flow turbines and bluff bodies at high blockage.
\end{abstract}

\maketitle

\section{Introduction}
\label{sec:intro}

Turbines deployed in river or tidal channels can convert the energy in the flow into mechanical power. The efficiency of energy conversion is influenced by the ``blockage ratio'' \citep{garrett_efficiency_2007} which quantifies the fraction of the channel cross-section $A_{\mathrm{channel}}$ occupied by the projected area of one or more turbines $A_\mathrm{{turbines}}$ as: 
\begin{equation}
    \beta = \frac{A_\mathrm{{turbines}}}{A_{\mathrm{channel}}} .
    \label{eq:blockage_general}
\end{equation}
As the blockage ratio increases, the turbines present greater resistance to the oncoming flow.
For a constant volumetric flow rate, this resistance is offset by confinement from the channel boundaries, resulting in an overall increase in flow through the turbine rotor and higher pressure drop.
As a consequence, turbines operating in a confined flow produce more power relative to the same turbines in unconfined conditions \citep{garrett_efficiency_2007, nishino_efficiency_2012, vogel_effect_2016, houlsby_power_2017}.

\begin{figure*}
    \centering
    \includegraphics[width=0.75\textwidth]{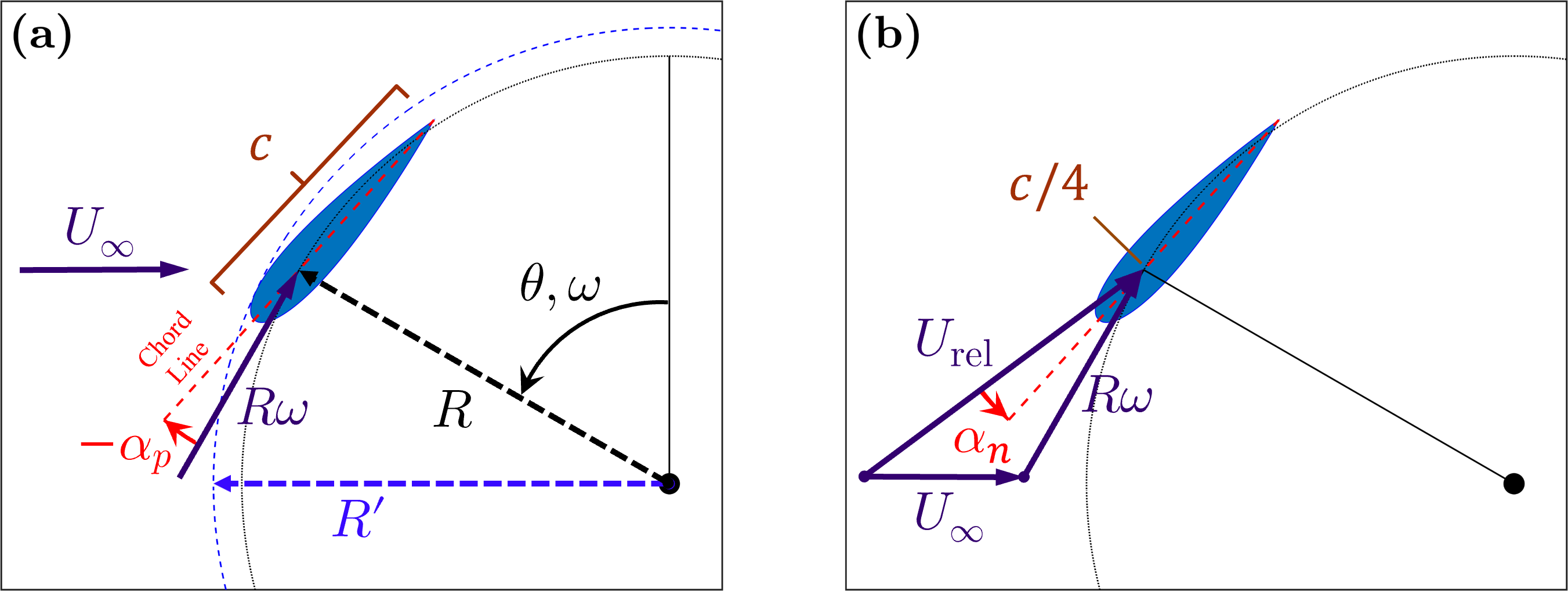}
    \caption{
    \textbf{(a)} Diagram of a cross-flow turbine blade with key geometric and kinematic quantities indicated.
    \textbf{(b)} The vector sum of the freestream velocity ($U_\infty$) and blade tangential velocity ($R_\omega$) yields a relative velocity on the blade ($U_{\mathrm{rel}}$) at an angle of attack $\alpha_n$, which results in lift and drag forces on the blade.
    The simplified representation in this figure does not consider effects of induction, flow curvature, and blockage that would alter the actual velocity incident on the blade and the corresponding location of the center of pressure (e.g., $c/4$ in (b)).}
    \label{fig:cftFlowVectors}
\end{figure*}

The definition of the blockage ratio in \Cref{eq:blockage_general}---which stems from linear momentum actuator disk theory---considers only the turbine projected area, and is agnostic to the particular turbine design that occupies this area.
While increasing $\beta$ increases the performance of all turbine designs, at a given $\beta$ the nature of performance augmentation can differ between turbine designs, and blockage effects have been quantified  experimentally \citep{whelan_freesurface_2009, ross_wind_2011, consul_blockage_2013, ross_experimental_2020, hunt_experimental_2024a} and numerically \citep{nishino_effects_2012, gauthier_impact_2016, kinsey_impact_2017, badshah_cfd_2019, zhang_analysis_2023} for various turbine types.
Cross-flow turbines---which have an axis of rotation perpendicular to the direction of inflow (\Cref{fig:cftFlowVectors}a)---have the potential to particularly benefit from confinement effects.
This is because, in river and tidal channels with a fundamentally rectangular form factor, the rectangular projected area of the cross-flow configuration allows relatively dense arrays to span a given channel.

Computational fluid dynamics (CFD) simulations of axial-flow turbines in confined flows have shown that optimal rotor design depends on the blockage ratio.
\citet{schluntz_effect_2015} utilized a blade element momentum (BEM) model embedded in a Reynolds-averaged Navier-Stokes (RANS) CFD solver to optimize the blade geometry of an axial-flow turbine for $\beta = 0\%$ (i.e., unconfined flow) to $31.4\%$.
The authors allowed the blade chord length and pitch to vary along the blade span, and primarily interpreted their results in the context of rotor solidity, a geometric parameter that represents the fraction of the turbine swept area that is occupied by the blades.
\citeauthor{schluntz_effect_2015} found that, as $\beta$ increased, the optimal rotor solidity increased and the optimal blade pitch decreased.
While the efficiency of all rotor geometries improved with increasing blockage ratio, the best performing rotor geometry at a given $\beta$ was the one optimized for that condition.
\citeauthor{schluntz_effect_2015} attributed these trends to the increased thrust on higher solidity rotors, which causes greater acceleration of the bypass flow and a correspondingly larger pressure drop across the rotor.
A similar trend was observed by \citet{abutunis_comprehensive_2022} in three-dimensional RANS simulations of three-bladed axial-flow turbines with various chord lengths and pitch angles at $\beta = 4.2\% - 62.5\%$. 
\citeauthor{abutunis_comprehensive_2022} observed that the efficiency of higher-solidity rotors (in this case, rotors with larger chord lengths) increased with $\beta$ at a faster rate than for rotors with lower solidity, and that this trend became more pronounced as the rotation rate increased.

While their rectangular projected area allows cross-flow turbines to achieve higher blockage ratios more easily than axial-flow turbines, they are comparatively understudied due to their inherently unsteady fluid mechanics.
Since the magnitude and direction of the relative velocity on the blades (\Cref{fig:cftFlowVectors}b) vary with blade position, the forces and torques on a cross-flow turbine vary periodically and the rapid changes in the angle of attack can produce dynamic stall\citep{mccroskey_phenomenon_1981, buchner_dynamic_2018, le_fouest_dynamic_2022, dave_analysis_2023}.
Additionally, the blades interact with the turbine wake during their downstream sweep, which influences the net performance over the rotational cycle \citep{snortland_influence_2025}.
The complexity of these dynamics limits the effectiveness of reduced-order models (e.g., Blade Element Momentum \citep{schluntz_effect_2015, vogel_blade_2018}) for the optimization of cross-flow turbines, and often necessitates narrower investigations of turbine performance through experiments and CFD simulations.

Cross-flow turbine performance is influenced by the rotor's geometry, and in this study we consider three key geometric parameters (\Cref{fig:cftFlowVectors}a).
First, the ``chord to radius ratio'' ($c/R$)---the ratio of the blade chord length ($c$) to the radial distance to the blade quarter-chord ($R$)---influences the flow curvature experienced by the blade.
As $c/R$ increases, the relative velocity on the blade increasingly varies along the chord line, which alters the apparent shape of the blade relative to the flow (``virtual camber'') as well as the angle of attack along the blade (``virtual incidence'')  \citep{migliore_effects_1980, takamatsu_study_1985, balduzzi_blade_2015, rainbird_influence_2015}.
Second, the ``preset pitch angle'' ($\alpha_p$)---the angle between the blade chord line and rotational tangent, defined at the quarter-chord---alters the range of angles of attack experienced by the turbine blades during rotation, which in turn influences dynamic stall during the blade's upstream sweep \citep{buchner_dynamic_2018, dave_analysis_2023, le_fouest_dynamic_2022}.
Prior work has shown that cross-flow turbines benefit from moderate toe-out (i.e. leading-edge out) preset pitch \citep{klimas_effects_1981, takamatsu_effects_1985, gosselin_parametric_2016, somoano_dead_2018, hunt_experimental_2024}, shown in \Cref{fig:cftFlowVectors}a as a negative $\alpha_p$.
Third, turbine performance is influenced by the number of blades ($N$).
In low-blockage flows (i.e., $\beta < 20\%$) \citep{li_effect_2015, araya_transition_2017, miller_solidity_2021, hunt_experimental_2024}, the maximum turbine efficiency and corresponding tip-speed ratio decrease with $N$, whereas the net thrust force on the rotor increases with $N$.
Additionally, the amplitude of cyclic forcing during a rotational cycle (i.e., the ratio of the peak to average forcing) decreases as blades are added \citep{li_effect_2016, hunt_experimental_2024}.
While $c/R$, $\alpha_p$, and $N$ each have individual effects, interactions between these geometric parameters also influence overall turbine performance.
For example, the optimal $\alpha_p$ has been shown to become more negative (i.e., more toe-out) as the chord-to-radius ratio increases \citep{takamatsu_experimental_1991, hunt_experimental_2024}.
Similarly, $c/R$ and $N$ are commonly combined to express the rotor solidity which, for a straight bladed cross-flow turbine, is given as
\begin{equation}
    \sigma = \frac{N}{2 \pi} \frac{c}{R} \ .
    \label{eq:solidity}
\end{equation}
While solidity can be representative of certain aspects of turbine performance (e.g., the tip-speed ratio corresponding to maximum performance \citep{rezaeiha_optimal_2018, hunt_experimental_2024}), it is often an incomplete descriptor due to the distinct hydrodynamic influences of $c/R$ and $N$ \citep{hunt_experimental_2024}.
An in-depth experimental investigation of the effects of $c/R$, $N$, $\alpha_p$, and $\sigma$ on cross-flow turbine performance at $\beta\!\approx\!11\%$ is provided by \citet{hunt_experimental_2024}.
In addition to the three parameters considered here, others can influence turbine performance.
These include the blade profile (e.g., symmetric or cambered) \citep{migliore_comparison_1983, takamatsu_experimental_1991, bianchini_design_2015, ouro_effect_2018, du_experimental_2019, chi_influence_2024}, blade shape (e.g., straight, canted, or helical) \citep{shiono_output_2002, armstrong_flow_2012, saini_review_2019}, aspect ratio \citep{bianchini_design_2015, li_effect_2017, hunt_effect_2020} and surface roughness \citep{howell_wind_2010, priegue_influence_2017}.

The effect of blockage on the optimal rotor geometry of a cross-flow turbine has seen limited investigation, with numerical simulations suggesting that, as for axial-flow turbines, the relationship between solidity and confinement influences performance.
\citet{goude_simulations_2014} simulated a cross-flow turbine at $\beta = 0-50\%$ using a two-dimensional vortex method, and varied the turbine solidity via the chord length.
The authors found that, at constant $\beta$, efficiency increased as solidity increased, and incremental changes in solidity yielded larger increases in efficiency at higher $\beta$.
\citet{kinsey_impact_2017} simulated both one-bladed and three-bladed cross-flow turbines at $\beta = 0\%$, $26\%$, and $\approx\!50\%$ using three-dimensional unsteady RANS, and found that the power and thrust coefficients of the three-bladed rotor increased more rapidly with blockage than those of the single-bladed rotor.
However, neither \citet{goude_simulations_2014} nor \citet{kinsey_impact_2017} tested a wide enough range of $\sigma$ to determine the optimal solidity at each $\beta$.
Additionally, the simulations by \citet{goude_simulations_2014} and \citet{kinsey_impact_2017} were not validated with experimental data at similar blockage ratios (likely due to scarce experimental performance data for cross-flow turbines at high blockage \citep{hunt_experimental_2024a}).

To the best of our knowledge, there have been no prior experimental investigations that parametrically examine how the optimal cross-flow turbine geometry varies with the blockage ratio, although some studies have varied individual geometric parameters at a constant, relatively high value of blockage.
For example, experimental work by \citet{mcadam_experimental_2013,mcadam_experimental_2013a} ($\beta = 47\%$) and \citet{takamatsu_study_1985,takamatsu_effects_1985} ($\beta = 75\%$) explored independent variation of $N$ and $\alpha_p$ for cross-flow turbines at high blockage.
\citet{mcadam_experimental_2013} observed a continual decrease in maximum efficiency with blade count for $N = 3 - 6$ at $\beta = 47\%$, and \citet{takamatsu_study_1985} found that two-bladed turbines outperformed four-bladed turbines at $\beta = 75\%$.
While these results suggest limitations to the benefits of increasing solidity at high blockage observed in simulation, they cannot provide insight into how the optimal rotor geometry changes as blockage varies since each study was performed at a single blockage ratio.
Furthermore, differences in experimental approach between the two studies preclude a synthesis of results.

Separately, several experimental studies have identified similarities between the dynamics of cross-flow turbines and bluff bodies (i.e., $\sigma \rightarrow$ 1). 
\citet{araya_transition_2017} found that the wake dynamics of a cross-flow turbine at $\beta = 20\%$ began to resemble those of a solid cylinder as both the rotation rate and the number of blades were increased.
To describe the composite geometric and kinematic effect, \citeauthor{araya_transition_2017} combined the ``static'' rotor solidity and tip-speed ratio into a ``dynamic solidity'' parameter and showed that the turbine wake developed bluff-body characteristics more rapidly at higher dynamic solidity.
Additionally, \citet{hunt_experimental_2024a} experimentally characterized the performance and near-wake flow field of a dual-rotor cross-flow turbine array at $\beta = 30\% - 55\%$, and showed that the array-average power and thrust coefficients were self-similar across the tested $\beta$ when scaled by the velocity of the fluid that bypasses the array.
This expanded on prior work by \citet{whelan_freesurface_2009} to describe turbine performance in confined flow in a manner inspired by the bluff-body theory of \citet{maskell_ec_theory_1963}.
Specifically, \citet{maskell_ec_theory_1963} hypothesized that the thrust on a bluff body responds to the accelerated flow that bypasses it.
Consequently, the self-similar performance observed by \citet{hunt_experimental_2024a} suggests that as the blockage ratio increases, cross-flow turbine dynamics resemble those of a bluff body.

Overall, while prior work suggests a relationship between confinement and cross-flow turbine geometry, the effects of $c/R$, $N$, $\alpha_p$ on performance have not been systematically explored as a function of $\beta$.
In addition, prior work has often framed geometric trends in terms of rotor solidity, which is a function of two parameters ($c/R$ and $N$) with distinct hydrodynamic effects \citep{hunt_experimental_2024}.
No prior studies---experimental or numerical---have examined how blockage affects the optimal preset pitch angle.
Additionally, while similarities between cross-flow turbine and bluff-body dynamics have been identified as a function of blade count \citep{araya_transition_2017} and blockage \citep{hunt_experimental_2024a}, this observation has not been generalized over a broad geometric and operating space. 
Such insights can be gained through experiments that examine how the individual and combined effects of $c/R$, $N$, and $\alpha_p$ evolve across a range of blockages. Here, the interplay between cross-flow turbine rotor geometry and confinement is explored with a two-rotor cross-flow turbine array.
Array performance is characterized for various geometric configurations at $\beta = 35\%$, $45\%$, and $55\%$, which represent the upper end of blockage ratios that are practically achievable in a tidal or river channel.
\Cref{sec:methods} describes the experimental methods employed, which are similar to those in prior work \citep{hunt_experimental_2024, hunt_experimental_2024a}.
\Cref{sec:results} presents the performance trends measured across the parameter space and compares these to prior results at lower blockage.
In \Cref{sec:discussion}, the hydrodynamic implications of these results are discussed and interpreted using a combination of two bluff body models: the concept of dynamic solidity introduced by \citet{araya_transition_2017} and a \citet{maskell_ec_theory_1963}-inspired linear momentum and bluff-body model.
The study concludes with a summary of key results and their implications for high-blockage turbine design in \Cref{sec:conclusion}. 
 
\section{Methods}
\label{sec:methods}


\subsection{Geometric Parameters}
\label{methods:geomParams}

\begin{table}[t]
    \centering
    \caption{Geometric parameters of the tested turbines. The average values of $R$ and $R'$ over all tested geometric configurations are shown, with the ranges of $R$ and $R'$ across the tested parameter space given in parentheses}
    \label{tab:geomParams}
    \begin{threeparttable}
        \begin{tabular}{@{}cc@{}}
            \toprule
            \textbf{Geometric Parameter} & \textbf{Value} \\ \midrule
            $R$ [cm] & 15.01 (14.73 - 15.08) \\
            $R'$ [cm] & 15.75 (15.31 - 16.16) \\
            $c$ [cm] & 5.57, 7.42, 9.28 \\ \midrule
            $c/R$ & 0.37, 0.49, 0.62 \\
            $\alpha_p$ [deg.] & 0, -2, -4, -6, -8, -10, -12 \\
            $N$ & 1, 2, 3, 4 \\ 
            $\sigma$ & 0.06 - 0.39 \\
            \bottomrule
        \end{tabular}
    \end{threeparttable}
\end{table}

\begin{figure*}[t]
    \centering
    \includegraphics[width=\textwidth]{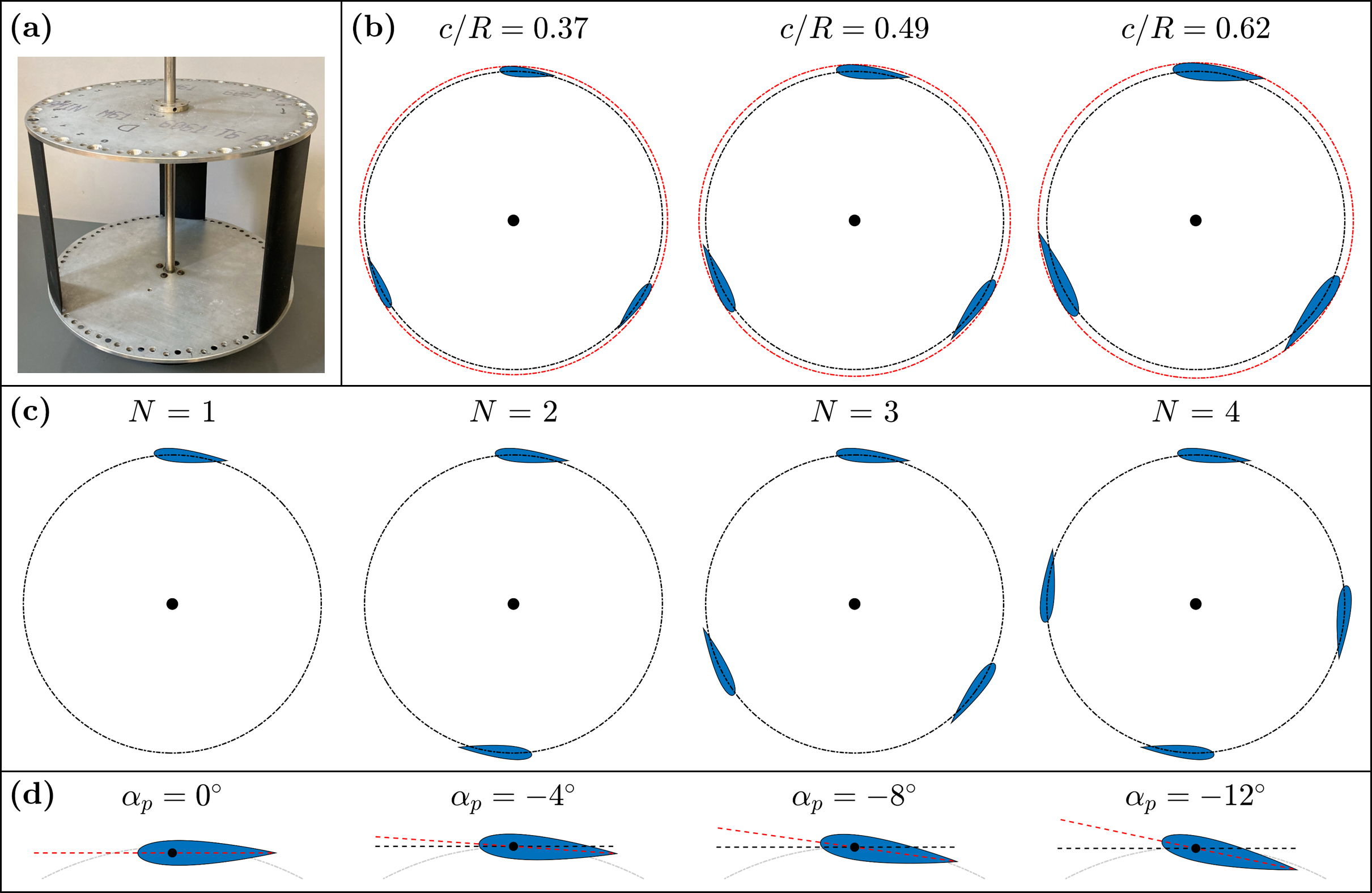}
    \caption{Partial visualization of experimental parameter space. \textbf{(a)} Image of an assembled rotor with $N = 3$, $c/R = 0.49$, and $\alpha_p = -6^{\circ}$. \textbf{(b)} Variation in $c/R$ for $N$ = 3 and $\alpha_p = -6^{\circ}$, with the quarter-chord radius $R$ shown in black and the resulting outermost swept radius $R'$ shown in red. \textbf{(c)} Variation in $N$ for $c/R = 0.49$ and $\alpha_p = -6^{\circ}$. \textbf{(d)} Variation in $\alpha_p$ for $c/R = 0.49$ with the chord line indicated in red and tangent line indicated in black.}
    \label{fig:geomShowcase}
\end{figure*}

The geometric parameter space explored in this work overlaps with the one explored at lower blockage by \citet{hunt_experimental_2024} and comprises three nominal chord-to-radius ratios ($c/R = 0.37$, $0.49$, and $0.62$), four blade counts ($N = 1$, $2$, $3$, and $4$) and seven preset pitch angles ($\alpha_p = 0^{\circ}$, $-2^{\circ}$, $-4^{\circ}$, $-6^{\circ}$, $-8^{\circ}$, $-10^{\circ}$, and $-12^{\circ}$).
Both $c/R$ and $\alpha_p$ are defined with respect to the quarter-chord point as in \cref{fig:cftFlowVectors}a).
The chosen values of $c/R$ and $N$ result in rotor solidities (\cref{eq:solidity}) ranging from 0.06 to 0.39.
These geometric parameters and the corresponding turbine dimensions are summarized in \Cref{tab:geomParams}.
At each $c/R$, five values of $\alpha_p$ are tested so as to both resolve the optimal preset pitch angle and characterize turbine performance at non-optimal pitch.
Relative to our prior work, a narrower range of $c/R$ is explored, but a wider range of $\alpha_p$ are considered at each $c/R$.
In total, $60$ unique combinations of $c/R$, $N$, and $\alpha_p$ are considered (a subset of which are visualized in \Cref{fig:geomShowcase}), with both turbines in the array sharing the same geometry in a given experiment. 
The blade span ($H$ = 21.5 cm) and foil profile (NACA 0018) are constant for all geometric configurations.
Additionally, to assess the limiting case of rotor solidity ($N \rightarrow \infty$), cylindrical shells with $H = 21.5$ cm and $R' = 15.75$ cm (i.e., the same outermost radius as the $c/R = 0.49$ turbines) were also tested.

All turbine blades are machined from 7075-T6 aluminum, anodized, and painted with an ultra-flat black spray paint.
The cylindrical shells are machined from 6061-T6 aluminum and painted as for the turbine blades.
As in our prior work\citep{hunt_experimental_2024}, the turbine blades are connected to the central driveshaft via circular end plates, which allows for different numbers of equally-spaced blades to be mounted at various preset pitch angles.
To economically explore the parameter space, the same sets of end plates are used for all blades.
However, for the smallest chord length ($c$ = 5.57 cm) to be compatible with the same end plates used for the larger chord lengths, the chord near the blade ends is gradually increased to 7.42 cm to accommodate the mounting hardware, while the middle $91\%$ of the blade span maintains the target chord length.
Additionally, because of the common end plates used, the radius of the outermost circle swept by the blades ($R'$ in \Cref{fig:cftFlowVectors}a) varies slightly for each combination of $c/R$ and $\alpha_{p}$.
For all end plates, the radial mounting position for a blade at each preset pitch angle is selected such that the $R'$ is constant for all preset pitch angles when using the mid-range chord length ($c$ = 7.42 cm).
Consequently, for a blade mounted at a given $\alpha_p$, while the quarter-chord position ($R$) and chord line orientation are constant across the chord lengths, $R'$ changes slightly due to the variation in foil thickness with $c$ (e.g., the red circles in \Cref{fig:geomShowcase}b).
However, these differences are relatively subtle.
Over the tested range of $c/R$ and $\alpha_p$, $R'$ for an individual configuration deviates by $<3\%$ from the average $R'$ given in \Cref{tab:geomParams}, which translates to deviations from the notional blockage ratios by $<2$ percentage points.
Similarly, the quarter-chord radius ($R$) varies slightly with $\alpha_p$ (e.g., $<2\%$ deviation from average value in \Cref{tab:geomParams}) to maintain constant $R'$ at $c/R = 0.49$.
A complete list of $R$, $R'$, and derived quantities for all geometric configurations is tabulated in \Cref{app:radVar}.
The treatment of the variation in the quarter-chord and outermost radii follows the approach in our prior work\citep{hunt_experimental_2024}, where $R$ is used to express all kinematic quantities (e.g., $\lambda$, $c/R$) and $R'$ is used to calculate the turbine projected area for the blockage and non-dimensional performance metrics.

\subsection{Experimental Set-Up}

\begin{figure}
        \centering
        \includegraphics[width=0.6\textwidth]{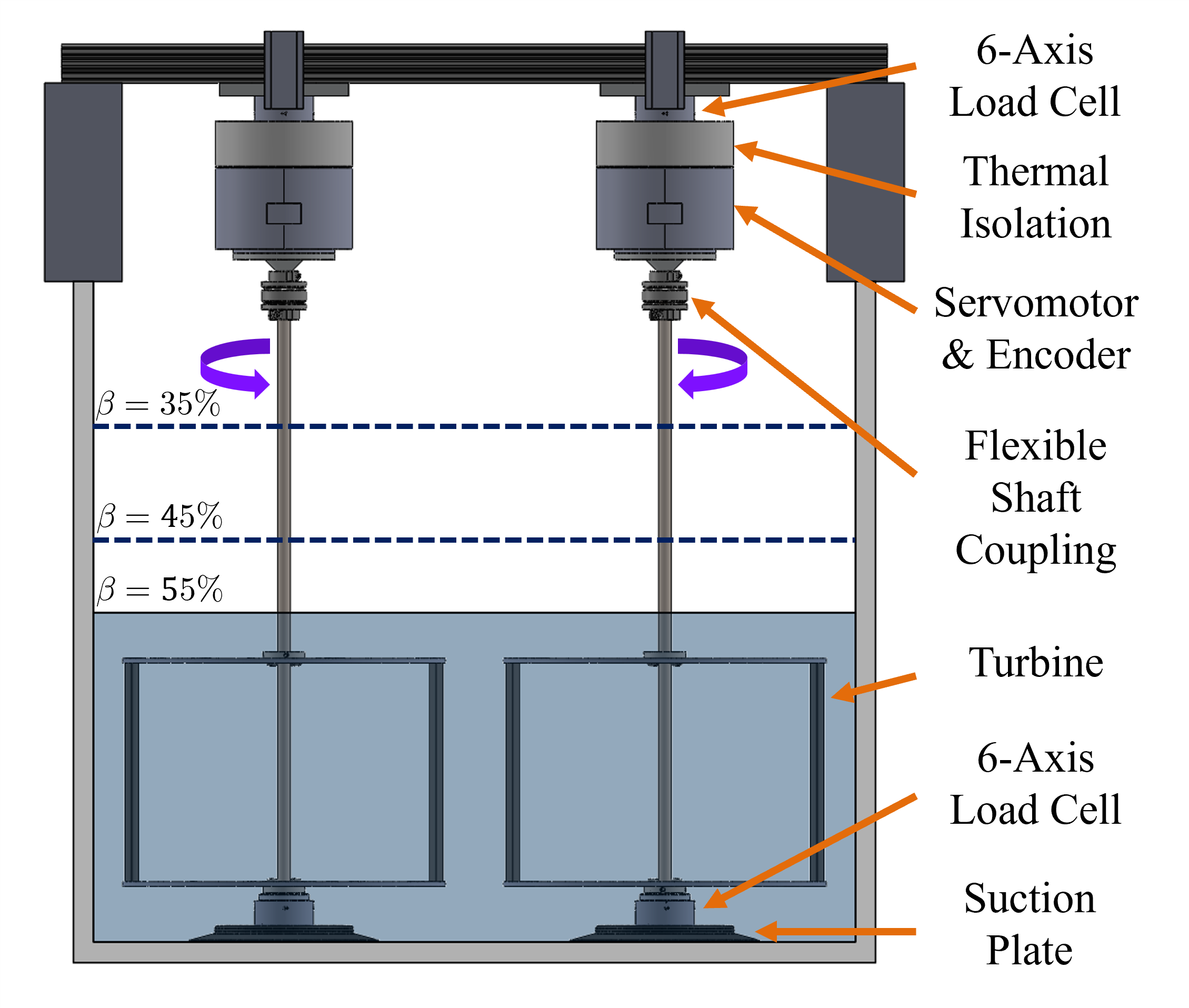}
        \caption{A rendering of the experimental setup as viewed from upstream, with key components annotated. The array shown is at $55\%$ blockage, and the dashed lines indicate the increasing water level for $45\%$ and $35\%$ blockage.}
        \label{fig:testRig}
\end{figure}

Experiments were conducted at the Alice C. Tyler recirculating water flume at the University of Washington.
The flume test section is 0.76 m wide and 4.88 m long, with a maximum fill depth of 0.60 m.
During experiments, the water temperature can be maintained between $10^{\circ} \mathrm{C}$ and $40^{\circ}\mathrm{C}$ using a heater and chiller.

The laboratory-scale array setup (\Cref{fig:testRig}) is as described in prior work \citep{hunt_experimental_2024a} and consists of two identical test-rigs.
The upper end of the central driveshaft of each turbine is connected via a flexible shaft coupling (Zero-Max SC040R) to a servomotor (Yaskawa SGMCS-05BC341), which regulates the rotation rate of each turbine to a prescribed value. This approach allows turbine performance to be characterized over the full range of operating conditions that result in net power generation \citep{polagye_comparison_2019}, but does not provide information about self-starting behavior for a given rotor geometry.
The instantaneous angular position of each turbine is measured using the servomotor encoder, from which the angular velocity is estimated.
The lower end of each turbine driveshaft sits in a bearing.
Forces and torques on each turbine are measured by a pair of six-axis load cells: an upper load cell (ATI Mini45-IP65) connected between the servomotor and a fixed crossbeam, and a lower load cell (ATI Mini45-IP68) connected between the bearing and a suction plate secured to the bottom of the flume.
Measurements from all load cells and servomotor encoders are collected synchronously at 1000 Hz using two National Instruments PCIe-6353 DAQs integrated with MATLAB.

The freestream velocity, $U_{\infty}$, is measured using an acoustic Doppler velocimeter (Nortek Vectrino Profiler).
$U_{\infty}$ is sampled at a point 5 turbine diameters upstream of the array centerline, positioned laterally in the center of the flume and positioned vertically at the blade midspan.
Velocity profiles at target inflow conditions without the array present indicate minimal vertical and lateral shear in the freestream (see supplemental information), such that a point measurement of $U_{\infty}$ is expected to be reasonably representative of rotor-averaged inflow.
Velocity measurements are collected at 16 Hz and despiked using the method of \citet{goring_despiking_2002}.
The water depth upstream of the array is measured at the center of the flume ${\sim}5.8$ turbine diameters upstream of the array centerline by an ultrasonic free surface transducer (Omega LVU32) sampling at 0.5 Hz.
The water temperature is measured using a temperature probe (Omega Ultra-Precise RTD) and maintained within $\pm 0.1^{\circ} \mathrm{C}$ of the target value during each experiment. Manufacturer-specified measurement uncertainties for all instruments are provided in supplemental information.

\subsection{Non-Dimensional Flow Parameters}
\label{methods:ndParams}

The blockage ratio for the laboratory-scale array is defined as
\begin{equation}
	\beta = \frac{A_{\mathrm{turbines}}}{A_{\mathrm{channel}}} = \frac{2HD'}{hw} ,
	\label{eq:arrayBlockage}
\end{equation}
\noindent where $D'$ = $2R'$, $h$ is the dynamic water depth, and $w$ is the channel width.
Each of the 60 geometric configurations is tested at blockage ratios of approximately $\beta = 35\%$, $45\%$, and $55\%$, resulting in 180 unique experiments.
The blockage ratio is varied by changing the water depth (\Cref{fig:testRig}).
To isolate the effects of geometry and blockage, other non-dimensional flow parameters that influence turbine performance are controlled.
For example, turbine performance also depends on the Reynolds number \citep{bachant_experimental_2016, rezaeiha_characterization_2018, miller_verticalaxis_2018, hunt_experimental_2024}, here defined with respect to the turbine diameter as
\begin{equation}
    Re_D =  \frac{U_{\infty} D}{\nu} \ \ ,
    \label{eq:ReD}
\end{equation}
where $D = 2R$ and $\nu$ is the kinematic viscosity.
Although turbine performance is invariant with the Reynolds number above a certain threshold \citep{bachant_experimental_2016, miller_verticalaxis_2018}, the relatively small size of the laboratory-scale rotors and the relatively low freestream velocities necessary to avoid ventilation (i.e., air entrainment in the rotor \citep{young_ventilation_2017}) at high blockage both constrain the achievable $Re_D$ to a transitional regime where turbine efficiency depends strongly on the Reynolds number \citep{miller_verticalaxis_2018}.
Although the nature of the blade boundary layer that governs the lift and drag forces on the turbine blades is driven by a ``local'' Reynolds number based on the relative velocity at the blade and the chord length, $Re_D$ is employed here for several reasons.
First, $Re_D$ is agnostic to blade geometry (which is intentionally varied through $c/R$).
Second, the relative velocity at the blade is not easily measured, and depends on properties of the turbine geometry and operating state.
The influence of variation in the local Reynolds number across experiments is discussed in \Cref{disc:ReynoldsEffects}.
Turbine performance also depends on the Froude number based on the channel depth \citep{consul_blockage_2013, vogel_effect_2016, hunt_effect_2020, ross_effects_2022}, given as
\begin{equation}
    Fr_h = \frac{U_{\infty}}{\sqrt{gh}} \ \ , 
    \label{eq:depthFroude}
\end{equation}
where $g$ is the acceleration due to gravity.
Therefore, to hold $Re_D$ and $Fr_h$ constant across the tested $\beta$, the freestream velocity and water temperature are varied as the water depth is varied \citep{hunt_effect_2020, hunt_experimental_2024a}.

\begin{table*}
    \centering
    \caption{Target and measured non-dimensional flow parameters and corresponding flume conditions for each blockage ratio. The range of time-averaged values (for a given tip-speed ratio set point) measured across all experiments are the italicized values given in parentheses. 
    }
    \label{tab:expMatrix}
    \setlength{\tabcolsep}{6pt}
    \renewcommand{\arraystretch}{0.75}
    \resizebox{\textwidth}{!}{
        \begin{tabular}{@{}cccc|cccc@{}}
        \toprule
        \multirow{2}{*}{\boldmath{$\beta \ [\%]$}} & \multirow{2}{*}{\boldmath{$Re_D \ [\times10^5]$}} & \multirow{2}{*}{\boldmath{$Fr_h$}} & \multirow{2}{*}{\boldmath{$s/h$}} & \multirow{2}{*}{\boldmath{$h$} \textbf{[m]}} & \multirow{2}{*}{\boldmath{$U_{\infty}$} \textbf{[m/s]}} & \multirow{2}{*}{\boldmath{$s$} \textbf{[m]}} & \multirow{2}{*}{\boldmath{$T \ [{^\circ}\mathrm{C}]$}} \\
         &  &  &  &  &  &  &  \\ \midrule
        \begin{tabular}[c]{@{}c@{}}35.0\\      \textit{(33.7 - 36.8)}\end{tabular} & \begin{tabular}[c]{@{}c@{}}1.62\\      \textit{(1.53 - 1.63)}\end{tabular} & \begin{tabular}[c]{@{}c@{}}0.219\\      \textit{(0.212 - 0.218)}\end{tabular} & \begin{tabular}[c]{@{}c@{}}0.470\\      \textit{(0.469 - 0.473)}\end{tabular} & \begin{tabular}[c]{@{}c@{}}0.509\\      \textit{(0.506 - 0.509)}\end{tabular} & \begin{tabular}[c]{@{}c@{}}0.489\\      \textit{(0.473 - 0.487)}\end{tabular} & \begin{tabular}[c]{@{}c@{}}0.241\\      \textit{(0.237 - 0.241)}\end{tabular} & $24.3 \pm 0.1$ \\ \midrule
        \begin{tabular}[c]{@{}c@{}}45.0\\      \textit{(43.1 - 47.2)}\end{tabular} & \begin{tabular}[c]{@{}c@{}}1.62\\      \textit{(1.50 - 1.61)}\end{tabular} & \begin{tabular}[c]{@{}c@{}}0.219\\      \textit{(0.199 - 0.217)}\end{tabular} & \begin{tabular}[c]{@{}c@{}}0.320\\      \textit{(0.319 - 0.328)}\end{tabular} & \begin{tabular}[c]{@{}c@{}}0.396\\      \textit{(0.394 - 0.399)}\end{tabular} & \begin{tabular}[c]{@{}c@{}}0.431\\      \textit{(0.394 - 0.426)}\end{tabular} & \begin{tabular}[c]{@{}c@{}}0.128\\      \textit{(0.126 - 0.131)}\end{tabular} & $30.1 \pm 0.1$ \\ \midrule
        \begin{tabular}[c]{@{}c@{}}55.0\\      \textit{(52.6 - 57.7)}\end{tabular} & \begin{tabular}[c]{@{}c@{}}1.62\\      \textit{(1.52 - 1.62)}\end{tabular} & \begin{tabular}[c]{@{}c@{}}0.219\\      \textit{(0.209 - 0.218)}\end{tabular} & \begin{tabular}[c]{@{}c@{}}0.170\\      \textit{(0.159 - 0.177)}\end{tabular} & \begin{tabular}[c]{@{}c@{}}0.324\\      \textit{(0.319 - 0.326)}\end{tabular} & \begin{tabular}[c]{@{}c@{}}0.390\\      \textit{(0.374 - 0.388)}\end{tabular} & \begin{tabular}[c]{@{}c@{}}0.056\\      \textit{(0.051 - 0.058)}\end{tabular} & $35.0 \pm 0.1$ \\ \bottomrule
        \end{tabular}
    }
\end{table*}

The target and measured values of $\beta$, $Re_D$, $Fr_h$, and corresponding flume conditions at each target $\beta$ are given in \Cref{tab:expMatrix}.
Across all experiments, the turbulence intensity was $< 2.2\%$.
The target diameter-based Reynolds number in this study ($\approx 1.6\!\times\!10^5$) facilitates comparison to prior work \citep{hunt_experimental_2024} that explored a similar geometric parameter space with a single turbine at $\beta \approx 11\%$ and the same $Re_D$, as well as prior work \citep{hunt_experimental_2024a} exploring the performance and near-wake flow field of the same array at $\beta = 30\%$ - $55\%$ and the same $Re_D$ and $Fr_h$.
Although the target $Fr_h$ in this study ($\approx0.21$) is roughly half that of our prior work at low blockage \citep{hunt_experimental_2024} ($0.41$), this does not preclude comparison of general trends between these studies because the effect of $\beta$ on performance is expected to outweigh that of $Fr_h$ \citep{ross_effects_2022}.
Although the influence of $Fr_h$ on turbine performance is expected to increase with the blockage ratio \citep{vogel_effect_2016}, the lower $Fr_h$ used here relative to prior work\citep{hunt_experimental_2024} may limit this effect.
While the target $Re_D$ and $Fr_h$ are the same for all experiments, we note that the normalized proximity of the array to the free surface (given as $s/h$, where $s$ is the distance between the free surface and the top of the turbine blades) is allowed to vary across the tested $\beta$.
While $s/h$ could be held constant by adjusting the vertical position of the array in the water column, it is instead maximized at each $\beta$ by operating the turbine low in the water column to reduce the risk of ventilation, which severely degrades performance \citep{birjandi_power_2013, young_ventilation_2017, hunt_experimental_2024a}.
Although changing $s/h$ can influence turbine performance even in the absence of ventilation \citep{birjandi_power_2013, kolekar_performance_2015, kolekar_blockage_2019, ross_effects_2022}, prior work with this array at identical flow conditions found that maximizing $s/h$ at similar $\beta$ has negligible effects on turbine performance \citep{hunt_experimental_2024a}.

While the same nominal inflow conditions are targeted for all experiments at a given blockage ratio, the actual values of $Re_D$, $Fr_h$, $\beta$, and $s/h$ vary due to a combination of turbine-flume interactions, differences between geometric configurations, and experimental variability.
As for previous experiments in this flume\citep{hunt_effect_2020, hunt_experimental_2023}, the flow resistance associated with the turbines generally elevates the water depth upstream of the array and reduces the freestream velocity.
This causes a corresponding decrease in the actual $\beta$, $Re_D$, and $Fr_h$, the nature of which varies with rotation rate, rotor geometry and the blockage ratio.
Additionally, independent of these interactions, small differences in $R$ and $R'$ with $c/R$ and $\alpha_p$ (\Cref{app:radVar}) contribute to variations in $\beta$ and $Re_D$ across the parameter space.
Given that these effects vary across the parameter space as a function of $\beta$ and the geometric parameters of interest, we did not attempt to adjust inflow speed and depth to precisely maintain the target parameter values.
As shown in \Cref{tab:expMatrix} the variation in time-averaged parameters across all experiments is relatively narrow, with further information about the ranges of time-averaged $\beta$, $Re_D$ and $Fr_h$ for individual experiments provided as supplementary material.

\subsection{Array Layout and Control}
\label{methods:arrayLayout}

\begin{figure}
        \centering
        \includegraphics[width=0.5\textwidth]{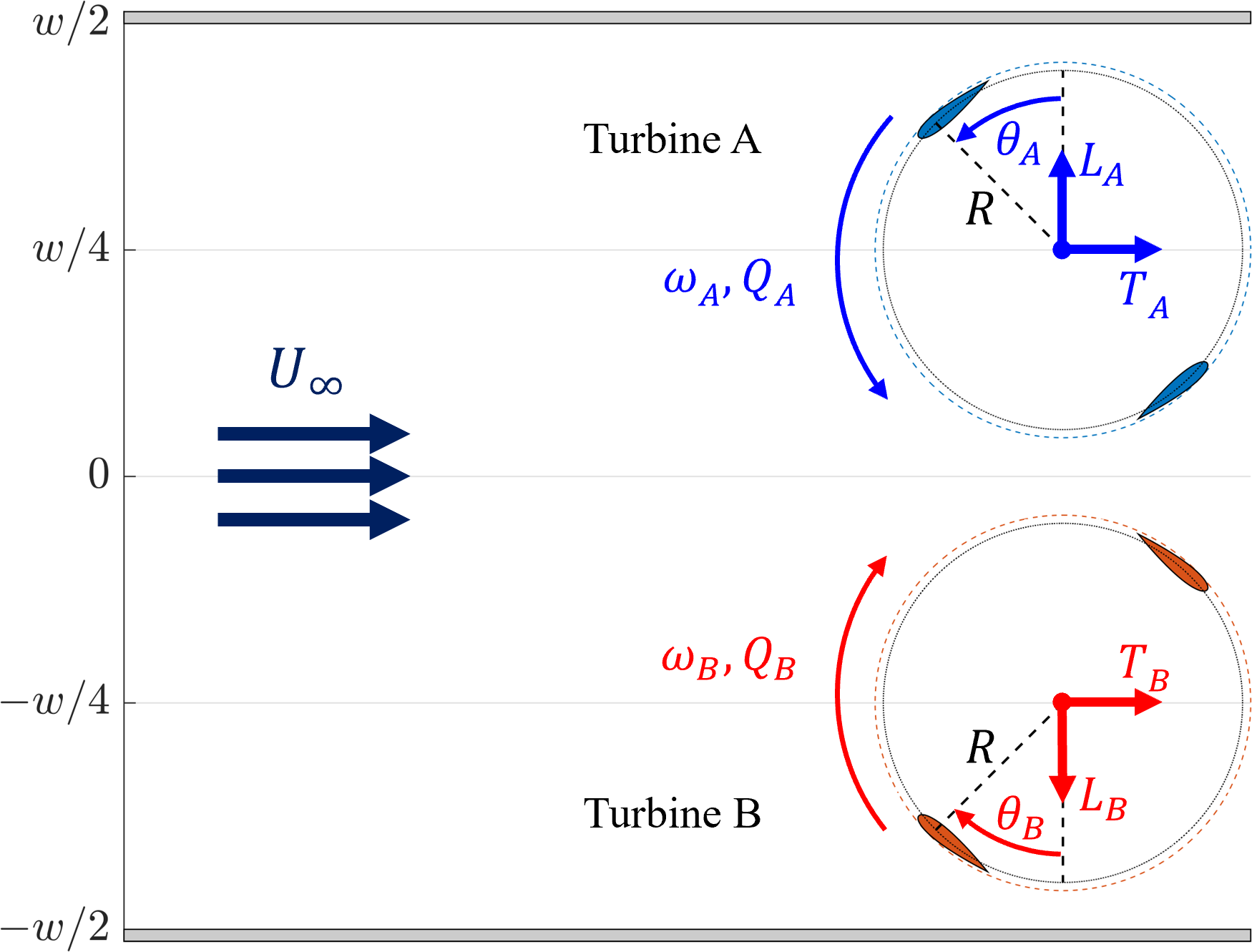}
        \caption{Overhead view of the array layout annotated with positive direction definitions for measured quantities. The outermost radii swept by the turbines ($R'$) are shown as the dashed lines.}
        \label{fig:arrayOverhead}
\end{figure}

An overhead view of the array layout is shown in \cref{fig:arrayOverhead}, in which the individual turbines in the array are designated as ``Turbine A'' and ``Turbine B''.
The array layout is identical to that in prior work by \citet{hunt_experimental_2024a}: the nominal center-to-center spacing between the turbines is ${\sim}1.2D'$, and the nominal blade-to-blade spacing between adjacent turbines (${\sim}0.22D'$) is twice the nominal wall-to-blade spacing such that the walls notionally correspond to symmetry planes in a larger array.
These spacings vary slightly across geometries due to small variations in $D'$ (\Cref{methods:geomParams}).
The turbines in the array are operated under a counter-rotating coordinated control scheme, wherein both turbines rotate in opposite directions at the same constant speed and with a constant angular phase offset, $\Delta \theta$, between them.
This control strategy is achieved by specifying the angular velocities of the rotors, which yields similar time-average performance to constant control torque \citep{polagye_comparison_2019}.
The turbines are counter-rotated such that the blades rotate toward the array centerline (\Cref{fig:arrayOverhead}), which has been found to enhance performance relative to alternative rotation schemes at lower blockage \citep{zanforlin_fluid_2016, scherl_optimization_2022, gauvin-tremblay_hydrokinetic_2022}.
The present experiments are limited to $\Delta \theta = 0^{\circ}$, an operating case in which the lateral forces and reaction torques for a pair of counter-rotating turbines are equal and opposite.
A closed-loop controller maintains $\Delta \theta$ to within $1^{\circ}$ of the target value at all rotation rates.

\subsection{Turbine performance metrics}
\label{methods:perfMetrics}

Array performance metrics are calculated from the measured quantities for each turbine.
The rotation rate of each turbine is non-dimensionalized as the tip-speed ratio, defined as the blade tangential velocity normalized by the freestream velocity:
\begin{equation}
    \lambda = \frac{\omega R}{U_{\infty}} \ \ .
    \label{eq:TSR}
\end{equation}
Data are collected at each tip-speed ratio for 45 seconds, and the time series is truncated to an integer number of turbine rotations for calculating performance metrics.

The efficiency (i.e., the coefficient of performance or power coefficient) of each turbine in the array is equal to the mechancial power produced by the turbine normalized by the kinetic power in the freestream flow that passes through the turbine's projected area
\begin{equation}
    C_{P}(t) = \frac{Q(t)\omega(t)}{\frac{1}{2}\rho \langle U_{\infty}^3(t) \rangle H D'} \ \ ,
    \label{eq:cp}
\end{equation}
\noindent where $Q(t)$ is the instantaneous hydrodynamic torque on the turbine and $\rho$ is the density of the working fluid.
Since measurements from the load cells and motor encoders are synchronized with each other but not with measurements of $U_{\infty}(t)$, the time-average of $U_{\infty}^3(t)$ is used to compute the instantaneous values of $C_P(t)$ at a given $\lambda$ set point as in prior work \citep{hunt_experimental_2024}.
The efficiency of each turbine is a function of the power produced by the blades and power losses due to parasitic torque on the support structures.
Because losses from the end plate support structures vary with rotation rate, direct comparisons between turbine geometries on the basis of $C_P$ would be convolved with differences in optimal rotation rate \citep{hunt_experimental_2024}.
Specifically, geometric configurations with the best blade-level performance at high $\lambda$ are penalized more steeply by parasitic torque on the end plates.
To facilitate comparisons across the parameter space, a blade-level efficiency for each turbine (i.e., the efficiency of the turbine blades in the absence of the support structures) is estimated via superposition following \citet{bachant_experimental_2016} and \citet{strom_impact_2018} as
\begin{equation}
    \label{eq:cpBlade}
        C_{P,\mathrm{blades}}(\beta, \lambda) \approx C_{P,\mathrm{turbine}}(\beta, \lambda) - C_{P,\mathrm{supports}}(\beta, \lambda) \ .
\end{equation}
The limitations of this approach---which assumes that secondary interactions between the blades and supports are negligible---for rotors at high blockage are discussed in \Cref{app:diskLoss}.
To summarize, the presence of appreciable secondary interactions in these experiments means that the values of blade-only efficiency reported are likely affected by the choice of support structure, although we do not expect trends across the parameter space to be significantly affected.

Structural loads on each turbine are characterized via the thrust and lateral force coefficients, given as
\begin{equation}
    C_{T}(t) = \frac{T(t)}{\frac{1}{2} \rho \langle U_{\infty}^2(t) \rangle H D'} \ \ ,
    \label{eq:cThrust}
\end{equation}
\begin{equation}
    C_{L}(t) = \frac{L(t)}{\frac{1}{2} \rho \langle U_{\infty}^2(t) \rangle H D'} \ \ ,
    \label{eq:cLat}
\end{equation}
\noindent where $T(t)$ and $L(t)$ are the instantaneous streamwise and cross-stream forces, respectively, on the turbine.
Since the thrust and lateral force on the support structures is small relative to that on the blades at high blockage (\Cref{app:diskLoss}), turbine-level $C_{T}$ and $C_{L}$ are approximately equal to the blade-level thrust and lateral force coefficients.
Since $C_{L}$ is not as strongly influenced by blockage as $C_P$ and $C_T$, discussion of the lateral force coefficients is deferred to supplementary material.

\subsection{Array Performance and Data Synthesis}

\begin{figure*}[t]
    \centering
    \includegraphics[width = 0.75\textwidth]{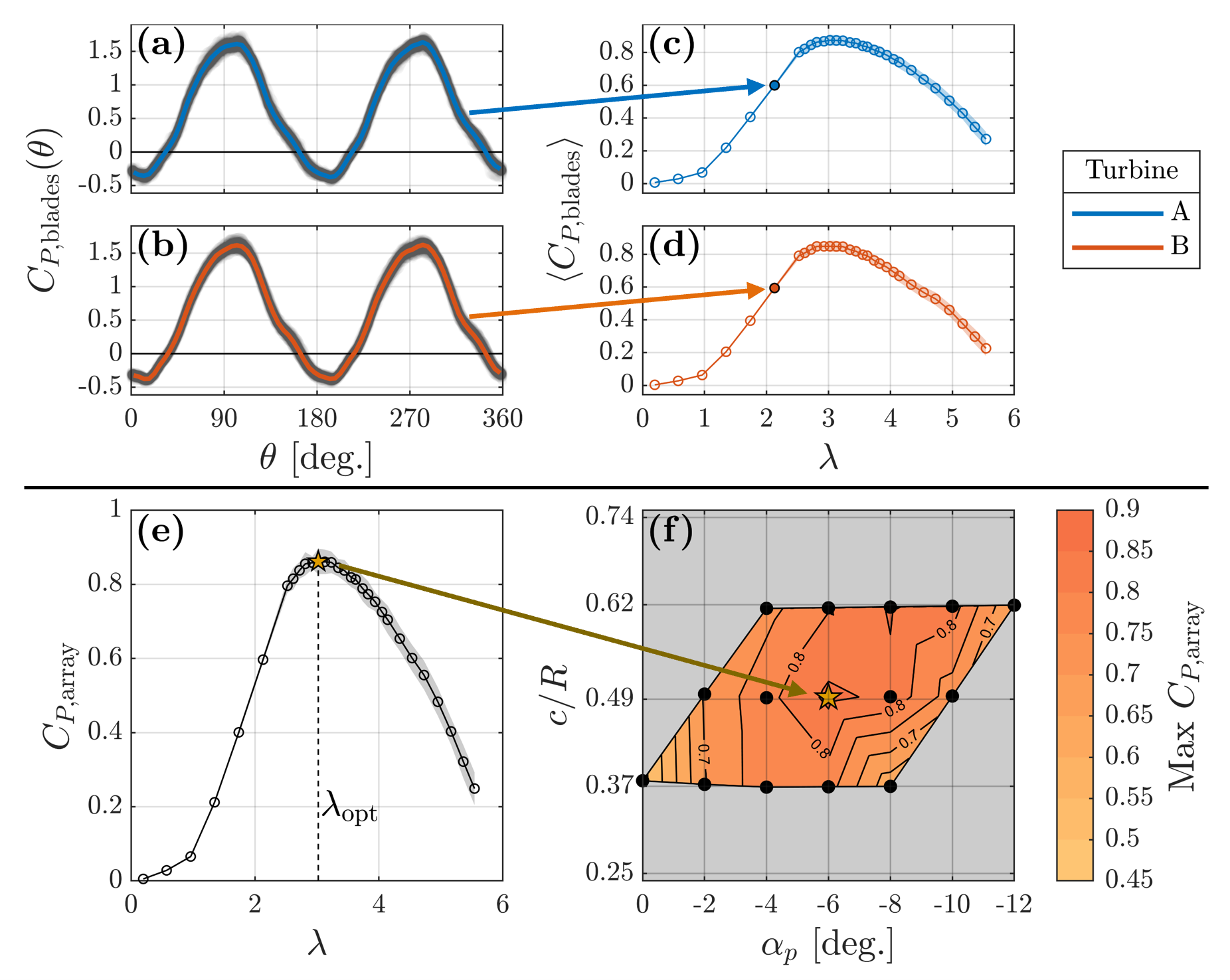}
    \caption{Construction of performance contours from the instantaneous blade-level efficiency for each geometric configuration as a function of $c/R$ and $\alpha_p$ at $\beta = 45\%$ and $N = 2$. \textbf{(a--b)} Instantaneous blade-level performance for each turbine during a rotational cycle at a given $\lambda$. The blue and orange curves indicate the median cycles for each turbine, whereas the gray region indicates the spread in the instantaneous values. \textbf{(c--d)} The resulting time-average blade-level efficiency as a function of $\lambda$ each turbine. The shaded regions (which are approximately the same width as the marker size) indicate $\pm1$ standard deviation in the cycle-averaged values measured at each $\lambda$. \textbf{(e)} The array-average blade-level efficiency (obtained as the average of the performance curves in (c--d)), with the maximum performance and corresponding $\lambda$ annotated. The shaded region indicates $\pm1$ standard deviation in the array-averaged cycle-averages at each $\lambda$ and is approximately the same size as the line width at lower $\lambda$. \textbf{(f)} Placement of the maximum performance point for this $c/R\!-\!\alpha_p$ combination on a contour map for all geometric configurations at $\beta = 45\%$ and $N = 2$.}
    \label{fig:dataSynthesis}
\end{figure*}

Since the turbines in this array are identical, array-average performance metrics are simply the average of the individual turbine performance metrics.
For example, the blade-level array-average efficiency, $C_{P,\mathrm{array}}$ is simply the average $C_{P,\mathrm{blades}}$ between Turbine A and Turbine B.
Similarly, the array-average thrust coefficient, $C_{T,\mathrm{array}}$, is the average $C_{T}$ between Turbine A and Turbine B.

To visualize trends across all experiments, results are aggregated and presented as contour maps of array-average performance metrics (e.g., maximum $C_{P,\mathrm{array}}$) as a function of rotor geometry and blockage ratio.
\Cref{fig:dataSynthesis} graphically outlines the procedure for generating a contour map of maximum $C_{P,\mathrm{array}}$ at one $\beta$ and $N$.
For a given combination of $c/R$ and $\alpha_p$, array performance is characterized under coordinated constant speed control (\Cref{methods:arrayLayout}) at a specified $\lambda$, at which the instantaneous performance of each turbine is measured (\Cref{fig:dataSynthesis}a--b).
The time-average blade-level efficiency for each turbine at this $\lambda$ is represented as a point on the respective characteristic performance curve (\Cref{fig:dataSynthesis}c--d), and this process is repeated at other $\lambda$ to obtain representative performance curves for both turbines.
The average of these performance curves is the array-average performance curve (\Cref{fig:dataSynthesis}e).
The tip-speed ratio associated with the maximum $C_{P,\mathrm{array}}$ for this geometric configuration is designated as $\lambda_{\mathrm{opt}}$, and this maximum efficiency is represented as a single point on a map of $c/R$ versus $\alpha_p$ for one $\beta\!-\!N$ combination (\Cref{fig:dataSynthesis}f).
The process in \Cref{fig:dataSynthesis} is repeated for each $c/R\!-\!\alpha_p$ combination tested at the same $\beta\!-\!N$ combination, and linear interpolation is used to visualize the contours of maximum efficiency within the convex hull spanned by the tested $c/R$ and $\alpha_p$.
A similar procedure is used to visualize the $C_{T,\mathrm{array}}$ associated with maximum $C_{P,\mathrm{array}}$.
In \Cref{sec:results}, the contours of array performance metrics at individual $\beta\!-\!N$ combinations are tessellated to visualize the trends in those performance metrics across the tested parameter space.

Prior uncertainty analysis for comparable experimental setups in this facility has been conducted by \citet{snortland_cycletocycle_2023} and \citet{hunt_experimental_2024}. Both studies followed the American Society of Mechanical Engineers' standard on test uncertainty \citep{american_society_of_mechanical_engineers_test_2005}, and showed that cycle-to-cycle variations in turbine performance (i.e., the shaded regions in \Cref{fig:dataSynthesis}a--e) are driven primarily by periodic fluctuations in the measured inflow velocity rather than instrument uncertainty.
The same conclusion is reached through formal uncertainty analysis of these experiments, as shown for a representative case in supplementary information.
Across all experiments, the median standard deviation of the cycle-average values of $C_{P,\mathrm{array}}$ at $\lambda_{\mathrm{opt}}$ is 0.023 ($\approx 3.3\%$ of the corresponding average value of $C_{P,\mathrm{array}}$ over all cycles), and the median standard deviation in the cycle-average values of $C_{T,\mathrm{array}}$ at $\lambda_{\mathrm{opt}}$ is 0.037 ($\approx 1.3\%$ of the corresponding average value of $C_{T,\mathrm{array}}$ over all cycles).
While cycle-to-cycle variation increases with $\lambda$ in all experiments, it is not significantly correlated with $\beta$ or rotor geometry.

\section{Results}
\label{sec:results}

This section presents the key trends observed across the tested parameter space, with an emphasis on the performance and thrust coefficients at the maximum efficiency point for each geometric configuration and blockage.
Hydrodynamic interpretation of these trends is provided in \Cref{sec:discussion}.

\begin{figure*}[t]
    \centering
    \includegraphics[width = \textwidth]{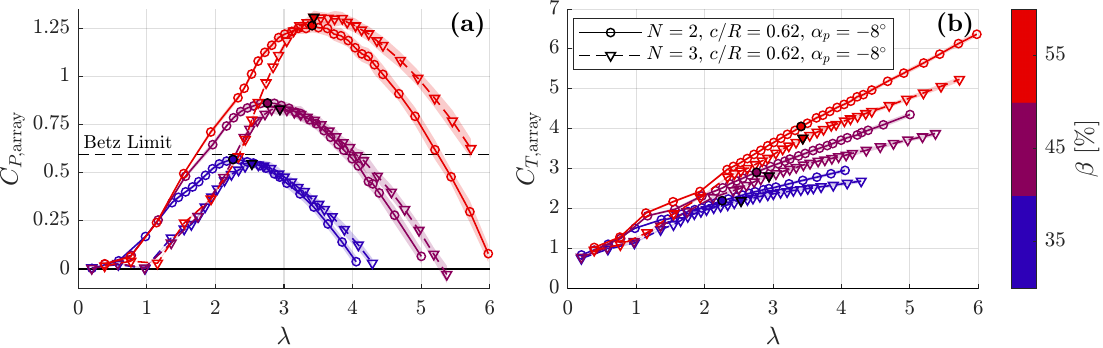}
    \caption{Time-average \textbf{(a)} $C_{P,\mathrm{array}}$ and \textbf{(b)} $C_{T,\mathrm{array}}$ as a function of $\lambda$ for two representative geometries. Data corresponding to maximum efficiency point are indicated by the filled markers. The shaded regions (which are approximately the same width as the plotted lines at lower $\lambda$) indicate $\pm1$ standard deviation in the cycle-average values.}
    \label{fig:representativePerfCurves}
\end{figure*}

Prior to presenting aggregate results, it is instructive to highlight the general effects of blockage on the characteristic array power and thrust curves, which are shown for two representative geometries in \Cref{fig:representativePerfCurves}.
As the blockage ratio increases, the maximum array efficiency increases, $\lambda_{\mathrm{opt}}$ increases, and the array produces power over a wider range of tip-speed ratios (\Cref{fig:representativePerfCurves}a).
Consistent with greater momentum changes, the array also experiences higher thrust as $\beta$ increases (\Cref{fig:representativePerfCurves}b).
Additionally, $C_{P,\mathrm{array}}$ and $C_{T,\mathrm{array}}$ are nearly invariant with $\beta$ up to a threshold value of $\lambda$, which is consistent with observations in prior studies \citep{consul_blockage_2013, kolekar_performance_2015, badshah_cfd_2019, hunt_experimental_2024a}.

In many cases, $C_{P,\mathrm{array}}$ exceeds the Betz limit for unconfined flow, and for several geometries at $\beta = 55\%$, $C_{P,\mathrm{array}}$ exceeds unity.
Such high efficiencies are not violations of energy conservation since the conventional definition of $C_P$ in \Cref{eq:cp} is referenced to the kinetic power in the flow that passes through the array projected area, and neglects the static head that can be harnessed in confined flows \citep{vogel_effect_2016, houlsby_power_2017}.
While alternative efficiency metrics that account for both the available static and dynamic head can be employed \citep{takamatsu_study_1985, mcadam_experimental_2013}, conventional velocity-based definitions of the power and thrust coefficients allow for comparison with prior studies at lower blockage.

\begin{figure*}[t]
    \centering
    \includegraphics[width=\textwidth]{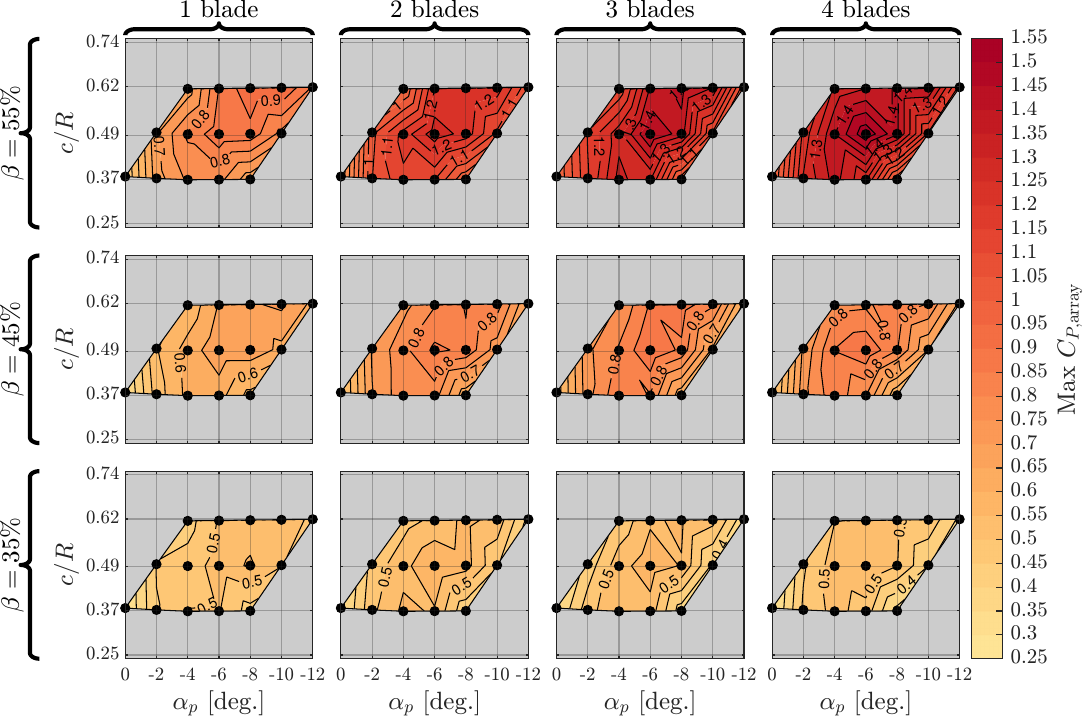}
    \caption{Contours of the maximum time-average array efficiency as a function of $c/R$ and $\alpha_p$ for each combination of $\beta$ and $N$ tested.}
    \label{fig:cpBladeHeatmap}
\end{figure*}

\begin{figure*}[t]
    \centering
    \includegraphics[width=\textwidth]{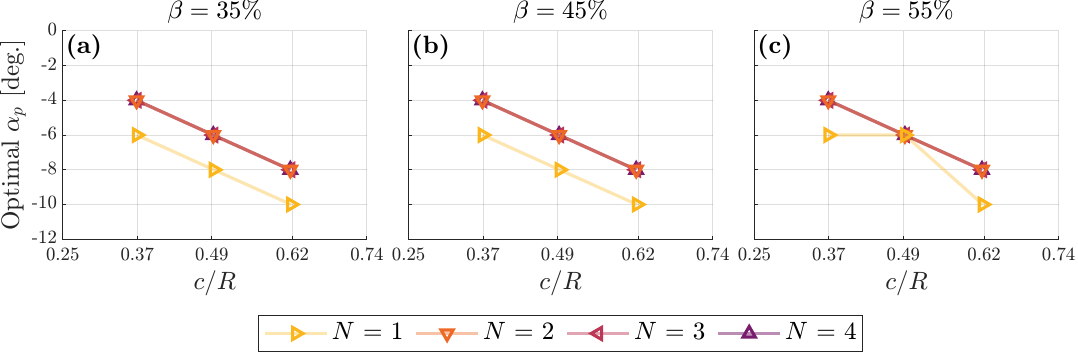}
    \caption{Optimal preset pitch angle as a function of $c/R$ and $N$ at each $\beta$. 
    Note that the $N=2$, $3$, and $4$ lines overlap in all cases.}
    \label{fig:optPitchLineplot}
\end{figure*}

Although the same general effects of blockage are observed for all rotor geometries, the geometric configuration of the rotor influences how these trends manifest.
As shown in \Cref{fig:representativePerfCurves}, the rotor geometry can change the rates at which $C_{P,\mathrm{array}}$, $C_{T,\mathrm{array}}$, and $\lambda_{\mathrm{opt}}$ increase with the blockage ratio, as well as the threshold value of $\lambda$ below which power and thrust are invariant with $\beta$.
To contextualize the effects of $c/R$, $N$, and $\alpha_p$ on array efficiency across the tested $\beta$, contours of maximum $C_{P,\mathrm{array}}$ across the entire parameter space are visualized in
\Cref{fig:cpBladeHeatmap}.
Although $\beta$ (i.e, moving between rows \Cref{fig:cpBladeHeatmap}) has the strongest overall effect, each geometric parameter influences $C_{P,\mathrm{array}}$ (i.e., there are no vertical or horizontal contours).
Trends between $c/R$, $\alpha_p$, and $C_{P,\mathrm{array}}$ largely follow those observed in prior studies of individual turbines at other blockages.
For example, the optimal preset pitch angle tends to become more negative (i.e., more toe-out) as the chord-to-radius ratio increases.
This relationship between $c/R$ and the optimal $\alpha_p$ is highlighted at each $\beta$ and $N$ in \Cref{fig:optPitchLineplot}, and is consistent with prior work at a single blockage ratio by \citet{hunt_experimental_2024} ($c/R = 0.25 - 0.74$; $\beta \approx 11\%$), and \citet{takamatsu_experimental_1991} ($c/R = 0.20 - 0.30$; $\beta = 75\%$).
The blade count has a secondary effect on the optimal preset pitch angle, and single-bladed turbines are generally observed to have more negative optimal $\alpha_p$ relative to that of two-, three-, and four-bladed turbines.
However, it is noted that, at a given $c/R$, single-bladed turbines often perform comparably well at the slightly more positive optimal $\alpha_p$ associated with $N=2-4$ (as seen explicitly in \cref{fig:optPitchLineplot}c for $c/R = 0.49$), such that the true optimal $\alpha_p$ for single-bladed turbines may be obscured by the $2^{\circ}$ resolution employed here.
Remarkably, trends in optimal $\alpha_p$ with $c/R$ and $N$ are largely invariant with $\beta$. 
Similarly, the best-performing $c/R$ does not change with $\beta$ across the tested parameter space.
Over the range of chord-to-radius ratios tested, $c/R = 0.49$ yields the greatest maximum $C_{P,\mathrm{array}}$ for $N=2-4$ at all $\beta$, with $c/R = 0.62$ performing comparably or better for $N=1$.

\begin{figure*}[t]
    \centering
    \includegraphics[width =\textwidth]{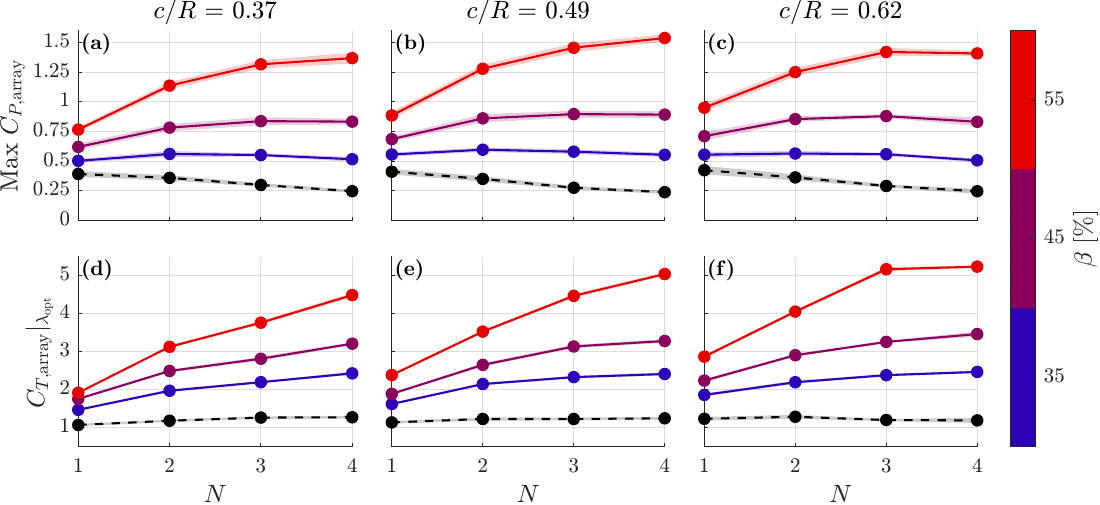}
    \caption{Maximum $C_{P,\mathrm{array}}$ at optimal $\alpha_p$ as a function of blade count and blockage ratio for \textbf{(a)} $c/R = 0.37$, \textbf{(b)} $0.49$, and \textbf{(c)} $0.62$.
    \textbf{(d)--(f)} $C_{T,\mathrm{array}}$ at maximum efficiency as a function of $N$ and $\beta$ for the same $c/R$ and $\alpha_p$.
    Note that the y-axis limits are substantially wider for thrust than efficiency.
    In all tiles, the dashed black lines represent the performance of a single turbine with the same geometric configurations at $
    \beta \approx 11\%$\citep{hunt_experimental_2024}.
    Shaded regions (which are no wider than the plotted marker size for all cases) indicate $\pm 1$ standard deviation in the cycle-average values.}
    \label{fig:blockageVsBlade}
\end{figure*}

In contrast, there is a strong interaction between the blade count and the blockage ratio, with the optimal number of blades tending to increase with $\beta$.
\Cref{fig:blockageVsBlade}a--c highlights how maximum $C_{P,\mathrm{array}}$ evolves with $\beta$ and $N$ for each $c/R$ at the corresponding optimal $\alpha_p$.
At $\beta$ = 35\%, $N=2$ yields the highest array-average efficiency for all $c/R$, whereas at $\beta = 45\%$, the best-performing blade count increases to $N=3$.
At $\beta = 55\%$, $N=4$ yields the highest $C_{P,\mathrm{array}}$ for $c/R = 0.37$ and $0.49$, whereas $N=3$ remains the most efficient blade count for $c/R = 0.62$.
For all $\beta$, the largest change in maximum efficiency occurs when $N$ increases from 1 to 2, with the subsequent addition of blades yielding smaller incremental changes in maximum efficiency.
These results are in agreement with the numerical results of \citet{kinsey_impact_2017} ($N=1$ and $N=3$; $\beta = 0-50\%$), and represent a notable reversal of the inverse relationship between blade count and maximum efficiency consistently observed at lower blockage ratios \citep{li_effect_2015, araya_transition_2017, miller_solidity_2021, hunt_experimental_2024} (e.g., the dashed black lines in \Cref{fig:blockageVsBlade}a--c, corresponding to the performance of a single turbine at $\beta = 11\%$ \citep{hunt_experimental_2024}) as well as in prior experimental studies at individual high blockage ratios \citep{takamatsu_effects_1985, mcadam_experimental_2013}.

\begin{figure*}[t]
    \centering
    \includegraphics[width = \textwidth]{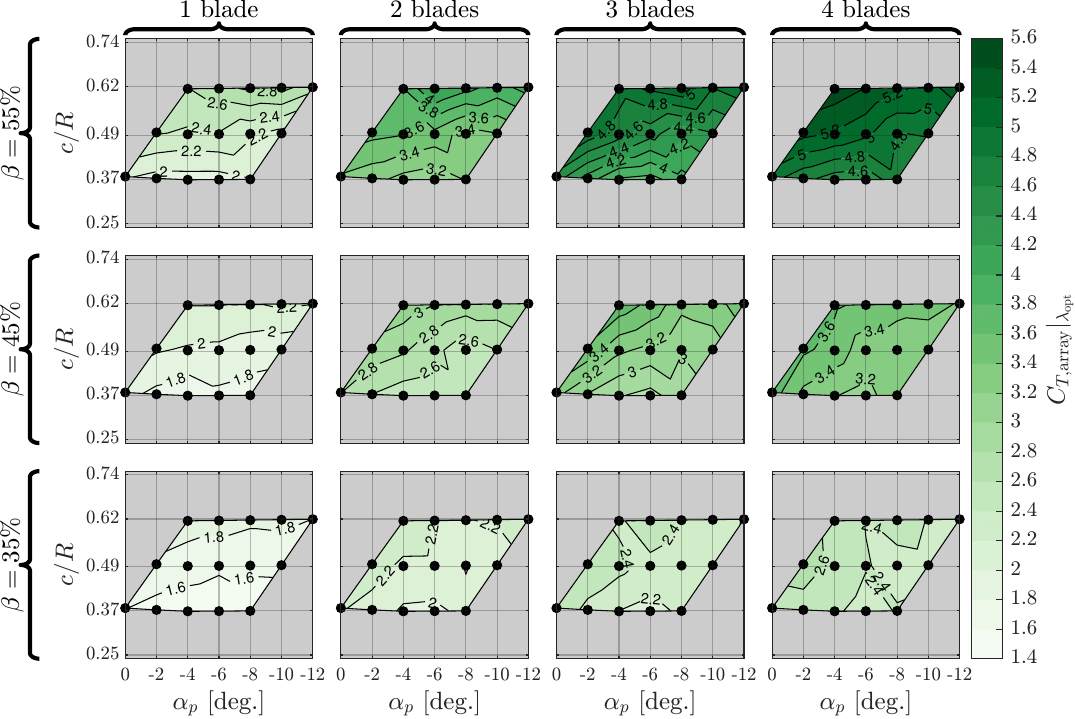}
    \caption{Contours of the time-average array thrust coefficient at maximum array performance as a function of $c/R$ and $\alpha_p$ at each $\beta$ and $N$ tested.}
    \label{fig:ctHeatmap}
\end{figure*}


A similar interaction between $N$ and $\beta$ is observed for the time-averaged array thrust coefficient at the optimal tip-speed ratio (i.e,. $C_{T,\mathrm{array}}\rvert_{\lambda_{\mathrm{opt}}}$), the contours of which are shown in \Cref{fig:ctHeatmap}.
For a given $c/R$ and $\alpha_p$, the array thrust coefficient at $\lambda_{\mathrm{opt}}$ increases as the number of blades increases, and this effect is amplified as the blockage ratio increases.
Relative to maximum $C_{P,\mathrm{array}}$, array thrust is more dependent on the blade count.
In a similar manner, larger $c/R$ increases $C_{T,\mathrm{array}}\rvert_{\lambda_{\mathrm{opt}}}$, and this effect is also augmented at higher $\beta$.
Of the tested parameters, $\alpha_p$ has the least influence on $C_{T,\mathrm{array}}\rvert_{\lambda_{\mathrm{opt}}}$, with array thrust slightly increasing for $\alpha_p$ less negative than optimal and decreasing for $\alpha_p$ more negative than optimal.
The influence of $N$ and $\beta$ on $C_{T,\mathrm{array}}\rvert_{\lambda_{\mathrm{opt}}}$ is shown at optimal $\alpha_p$ for each $c/R$ in \Cref{fig:blockageVsBlade}d--f, and highlights how the incremental increase in $C_{T,\mathrm{array}}\rvert_{\lambda_{\mathrm{opt}}}$ with additional blades is generally amplified as $\beta$ increases.
In contrast, for a single turbine at $\beta = 11\%$, \citet{hunt_experimental_2024} did not observe a significant change in $C_T$ at $\lambda_{\mathrm{opt}}$ with $N$ for these $c/R$.
This is because increasing $N$ yields both higher $C_T$ at a given $\lambda$ and a reduction in $\lambda_{\mathrm{opt}}$ \citep{li_effect_2015,hunt_experimental_2024}, and at $\beta=11\%$ these effects offset each other to result in relatively constant $C_T\rvert_{\lambda_{\mathrm{opt}}}$.
At higher blockage, the plateauing of $C_{T,\mathrm{array}}\rvert_{\lambda_{\mathrm{opt}}}$ at higher blade counts (e.g., \Cref{fig:blockageVsBlade}f) may be attributable to a similar effect.

\begin{figure*}[t]
    \centering
    \includegraphics[width = \textwidth]{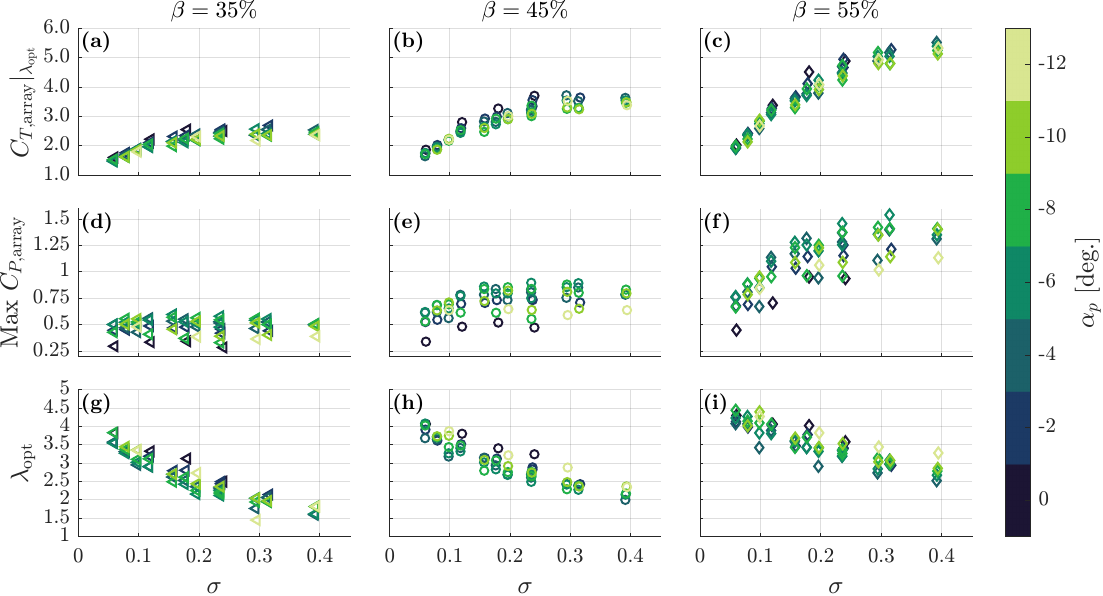}
    \caption{\textbf{(a)--(c)} $C_{T,\mathrm{array}}$ at max efficiency, \textbf{(d)--(f)} maximum $C_{P,\mathrm{array}}$, and \textbf{(g)--(i)} optimal tip-speed ratio as a function of solidity at each blockage.}
    \label{fig:solidityTrends}
\end{figure*}

        %
Since the array-average thrust at $\lambda_{\mathrm{opt}}$ increases with both the number of blades and the chord-to-radius ratio, their net effect might be well-represented by solidity, which combines these two geometric parameters (\Cref{eq:solidity}).
To investigate this, $C_{T,\mathrm{array}}\rvert_{\lambda_{\mathrm{opt}}}$ is regressed against solidity in \Cref{fig:solidityTrends}a--c.
A strong correlation between $\sigma$ and the thrust coefficient at $\lambda_{\mathrm{opt}}$ is observed at each blockage ratio, indicating that arrays with similar solidity have similar thrust coefficients at $\lambda_{\mathrm{opt}}$, regardless of the particular $N$ and $c/R$ employed.
Since confinement affects the flow through and around the turbine through the channel blockage ratio and the thrust coefficient \citep{garrett_efficiency_2007, houlsby_power_2017}, this suggests that the flow fields in the vicinity of rotors with similar solidities are similarly affected by blockage at $\lambda_{\mathrm{opt}}$.
$C_{T,\mathrm{array}}\rvert_{\lambda_{\mathrm{opt}}}$ increases more rapidly with $\sigma$ at higher $\beta$, but gradually plateaus at higher solidity in a manner similar to the results in \Cref{fig:blockageVsBlade}d--f.
At $\beta = 35\%$ and $\beta = 45\%$, $C_{T,\mathrm{array}}\rvert_{\lambda_{\mathrm{opt}}}$ is insensitive to solidity at the upper end of the tested $\sigma$, whereas this threshold is not reached for the tested $\sigma$ at $\beta = 55\%$.
Relative to solidity and the blockage ratio, $\alpha_p$ has a limited influence on $C_{T,\mathrm{array}}\rvert_{\lambda_{\mathrm{opt}}}$ at a given $\sigma$.

In contrast, solidity is marginally descriptive of trends in maximum array efficiency (\Cref{fig:solidityTrends}d--f).
As $\beta$ increases, the highest efficiencies are generally associated with higher solidity, which is in agreement with the results of prior numerical studies of cross-flow  \citep{goude_simulations_2014, kinsey_impact_2017} and axial-flow turbines \citep{schluntz_effect_2015, abutunis_comprehensive_2022} at high blockage.
At $\beta = 35\%$ (\Cref{fig:solidityTrends}d), maximum $C_{P,\mathrm{array}}$ is relatively insensitive to solidity.
However, as $\beta$ increases (\Cref{fig:solidityTrends}e,f), maximum $C_{P,\mathrm{array}}$ tends to increase more rapidly with solidity in a manner similar to $C_{T,\mathrm{array}}\rvert_{\lambda_{\mathrm{opt}}}$.
Because the optimal preset pitch angle is primarily influenced by one component of solidity ($c/R$; \Cref{fig:optPitchLineplot}), maximum $C_{P,\mathrm{array}}$ varies significantly with $\alpha_p$ at constant $\sigma$.
Additionally, $c/R$ likely influences $C_{P,\mathrm{array}}$ through unique hydrodynamic mechanisms that are not associated with $N$, as discussed in \Cref{disc:cToR}.
Consequently, relative to $C_{T,\mathrm{array}}\rvert_{\lambda_{\mathrm{opt}}}$, maximum $C_{P,\mathrm{array}}$ is more sensitive to the particular configuration of $c/R$, $N$, and $\alpha_p$, such that higher $\sigma$ is not always associated with higher maximum efficiency. 

Although solidity is an incomplete descriptor of the maximum array efficiency, it is a good predictor of the tip-speed ratio at which maximum efficiency occurs ($\lambda_{\mathrm{opt}}$).
In a manner similar to that observed in prior low-blockage experimental work by \citet{hunt_experimental_2024} and numerical work by \citet{rezaeiha_optimal_2018}, $\lambda_{\mathrm{opt}}$ decreases with $\sigma$ across the tested blockages (\Cref{fig:solidityTrends}g--i).
Compared to our prior work \citep{hunt_experimental_2024}, there is somewhat greater spread in $\lambda_{\mathrm{opt}}$ at each $\sigma$.
This is likely a consequence of the broader range of preset pitch angles tested at each $c/R$ in this study, and the corresponding larger shifts in $\lambda_{\mathrm{opt}}$ that occur further from the optimal value of $\alpha_p$.
While \Cref{fig:solidityTrends} focuses solely on the relationship between array performance and solidity at $\lambda_{\mathrm{opt}}$, the extension of the observed trends to other operating conditions is explored in \Cref{disc:dynamicSolidity}.

\section{Discussion}
\label{sec:discussion}

\subsection{Interplay between the Reynolds Number and Confinement}
\label{disc:ReynoldsEffects}

Before discussing the hydrodynamics underpinning the results, it is important to acknowledge the interplay between confinement effects and the local Reynolds number.
Although the target Reynolds number with respect to the rotor diameter and freestream velocity ($Re_D$, \Cref{eq:ReD}) is held approximately constant, the lift and drag forces on the turbine blades depend on Reynolds number local to the blade, given as
\begin{equation}
    Re_c =\frac{U_{\mathrm{rel}}c}{\nu} \ ,
    \label{eq:ReC}
\end{equation}
where $U_{\mathrm{rel}}$ is the relative velocity on the blade (\Cref{fig:cftFlowVectors}b).
In confined flow, the presence of channel boundaries increases flow through the array relative to an equivalent unconfined array \citep{garrett_efficiency_2007, houlsby_power_2017}.
For a given geometry and tip-speed ratio, this confinement-driven increase of the near-blade inflow velocity elevates $Re_c$.
Consequently, as noted by \citet{ross_effects_2022}, for a given rotor geometry and rotation rate, an increase in blockage effects is inextricably linked to an increase in $Re_c$.
This influences the blade boundary layer for experiments conducted in a transitional Reynolds number regime and, consequently, we would expect the blockage ratio to have less effect on performance in a Reynolds-independent regime. 

The geometric parameters considered in this study are also expected to influence $Re_c$ through various---and often competing---mechanisms.
For example, increasing $c/R$ increases $Re_c$ via the length scale $c$.
Through solidity, $c/R$ and $N$ are expected to increase induction \citep{rezaeiha_optimal_2018, hunt_experimental_2024} (which reduces $U_\mathrm{rel}$), but increases thrust (\Cref{fig:solidityTrends}) and thus increases flow through the rotors (which increases $U_\mathrm{rel}$).
Since a measurement of $U_{\mathrm{rel}}$ is unavailable, $Re_c$ is indeterminate and non-constant across the tested parameter space.

\subsection{Effects of Blade Count}
\label{disc:bladeCount}

The number of blades changes the resistance the turbines apply to the flow.
As $N$ increases, the rotor solidity and thrust coefficient increase (\Cref{fig:ctHeatmap}, \Cref{fig:solidityTrends}a--c), corresponding to a larger pressure drop across each turbine. In unconfined or low-blockage flows, the elevated resistance causes flow to divert around the rotor, such that increasing $N$ typically decreases maximum efficiency \citep{li_effect_2015, araya_transition_2017, rezaeiha_optimal_2018, miller_solidity_2021, hunt_experimental_2024}.
In confined flow, the constriction from the channel boundaries partially offsets this \citep{garrett_efficiency_2007,houlsby_power_2017}, allowing rotors with more blades to generate more power as blockage increases (\Cref{fig:cpBladeHeatmap}). 
While linear momentum theory establishes a direct connection between the pressure drop and $C_{T,\mathrm{array}}$, the manner in which the pressure drop influences $C_{P,\mathrm{array}}$ depends on rotor geometry and near-blade hydrodynamics which are not accounted for in linear momentum theory.
Additionally, the blade count likely influences turbine performance by changing the nature of blade-to-blade interactions, both within an individual rotor and between turbines in the array.
However, this mechanism is not explored further in this study.

For the range of $N$ considered in these experiments, the optimal blade count generally increases with the blockage ratio (e.g., \Cref{fig:blockageVsBlade}a--c).
We hypothesize that the optimal blade count likely balances the benefits of a larger pressure drop across the turbines (which increases for larger $N$) and the penalty of a lower mass flow rate through the rotors (which decreases for larger $N$), as suggested by the numerical results of \citet{schluntz_effect_2015}.
As $N \rightarrow \infty$, more flow will be diverted around the array rather than through the rotors, reducing maximum $C_{P,\mathrm{array}}$ and causing $C_{T,\mathrm{array}}$ to tend toward a steady-state value at a given $\lambda$.
Since, in these experiments, $C_{P,\mathrm{array}}$ does not vary significantly for $N = 1-4$ at $\beta = 35\%$, this particular blockage ratio may represent an inflection point in the effect of blade count between the low and high blockage regimes.

\begin{figure*}[t]
    \centering
    \includegraphics[width = \textwidth]{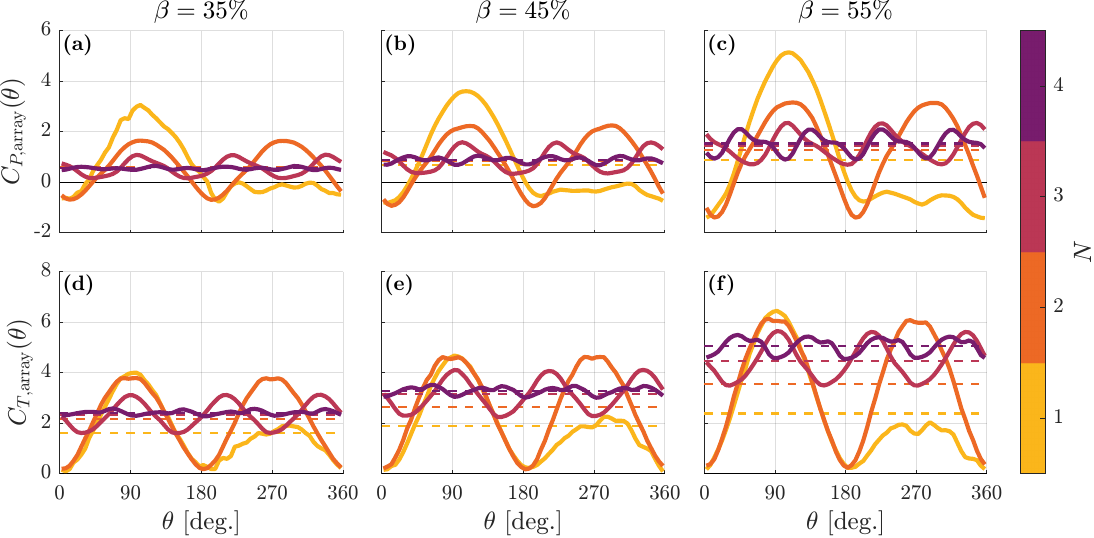}
    \caption{Phase-median \textbf{(a)--(c)} $C_{P,\mathrm{array}}$ and \textbf{(d)--(f)} $C_{T,\mathrm{array}}$ over a rotational cycle at $\lambda_{\mathrm{opt}}$ as a function of blade count and blockage for $c/R = 0.49$ and optimal $\alpha_p$. The dashed lines represent the time-average $C_{P,\mathrm{array}}$ and $C_{T,\mathrm{array}}$ over the median cycle.}
    \label{fig:bladePhaseTrends}
\end{figure*}

In addition to influencing the time-average array power and thrust, the number of blades influences the variation in turbine loading over the course of a rotational cycle.
These trends are explored at $\lambda_{\mathrm{opt}}$ for representative geometries with $c/R = 0.49$, optimal $\alpha_p$, and various $N$ across the tested $\beta$ in \Cref{fig:bladePhaseTrends}.
Although the cycle-average power and thrust coefficients increase with both $\beta$ and $N$, at a given $\beta$ the amplitudes of cyclic power (\Cref{fig:bladePhaseTrends}a--c) and thrust fluctuations (\Cref{fig:bladePhaseTrends}d--f) are reduced as $N$ increases.
In other words, although increasing the blade count increases the average force on the turbines, it reduces the phase-maximum forces experienced by the turbines.
This characteristic of the blade count is also present in unconfined and low-blockage flows \citep{li_effect_2016, hunt_experimental_2024}.
Since structural and power take-off requirements are driven moreso by the peak forces and torques than the time-averages, turbines with more blades may have an economic advantage in high blockage flows.
This benefit is especially highlighted in these experiments since there is only a limited reduction in maximum efficiency if one or two blades more than the optimal $N$ are used (e.g., \Cref{fig:bladePhaseTrends}a--c).

\subsection{Effects of Chord-to-Radius Ratio}
\label{disc:cToR}

Like the blade count, the chord-to-radius ratio changes the rotor solidity, and thus the resistance the turbines apply to the flow and the blockage effects experienced.
Therefore, an increase in $c/R$ increases the thrust on the array (\Cref{fig:ctHeatmap}), and the optimal $c/R$ is likely influenced by a tradeoff between a higher pressure drop across the rotors and reduced mass flow rate through the rotors.
However, while the relationship between $c/R$ and $C_{T,\mathrm{array}}\rvert_{\lambda_{\mathrm{opt}}}$ is well described by this interaction between solidity and blockage (\Cref{fig:solidityTrends}a--c), the influence of $c/R$ on $C_{P,\mathrm{array}}$ cannot be completely attributed to $\sigma$ and $\beta$ since $c/R$ simultaneously exerts several unique hydrodynamic influences that $N$ does not.
First, while $N$ and $c/R$ are both expected to influence the local Reynolds number at the blade ($Re_c$) through changes to the relative velocity from flow induction and blockage effects (\Cref{disc:ReynoldsEffects}), $c/R$ also directly changes $Re_c$ through the length scale $c$.
Second, the optimal preset pitch angle depends primarily on $c/R$ (\Cref{fig:optPitchLineplot}), such that a change in $c/R$ at constant $\alpha_p$ inherently convolves the effects of these two parameters.
Third, an increase in $c/R$ increases the flow curvature effects experienced by the blades \citep{migliore_flow_1980, takamatsu_study_1985, balduzzi_blade_2015}.

While the interplay between $c/R$ and $\alpha_p$ on $C_{P,\mathrm{array}}$ is explored in this study, isolating the contributions of $Re_c$ and flow curvature effects requires further investigation.
Furthermore, trends in $C_{P,\mathrm{array}}$ with $c/R$ may evolve as $Re_D$ increases, since, at lower blockage ratios, the optimal $c/R$ has been shown to decrease with the Reynolds number \citep{bianchini_design_2015, hunt_experimental_2024}.
Therefore, the optimal $c/R$ for arrays operating at both high $Re_D$ and high $\beta$ may be that which balances the confinement effects (facilitated by higher solidity) with the favorable hydrodynamics associated with lower $c/R$ at higher Reynolds numbers.

\subsection{Effects of Preset Pitch Angle}

The preset pitch angle influences array performance by altering the range of angles of attack experienced by the blades throughout a rotational cycle.
Although the actual angle of attack on the blades is unknown, a simplified model that neglects the influence of the rotor on the near-blade inflow velocities predicts that a negative preset pitch angle reduces the maximum magnitude of the angle of attack on the blade during the upstream sweep ($0^{\circ} \leq \theta < 180^{\circ}$) \citep{hunt_experimental_2024,snortland_influence_2025}.
Consequently, the preset pitch angle is expected to alter the characteristics of dynamic stall and flow separation during the upstream sweep.
In the downstream sweep, this simplified model is not descriptive of the flow field and the effect of the preset pitch angle on the angle of attack is ambiguous.
Prior experimental work by \citet{hunt_experimental_2024} at $\beta = 11\%$ has shown that the optimal preset pitch angle is that which balances power production during the upstream sweep with power consumption during the downstream sweep.
Given that the trends in $\alpha_p$ observed here parallel those of \citet{hunt_experimental_2024}, these hydrodynamic mechanisms are likely unchanged with blockage.
Since the optimal value of $\alpha_p$ depends primarily on $c/R$, has a secondary dependence on $N$ for single-bladed turbines (\Cref{fig:optPitchLineplot}), and is independent of $\beta$, $\alpha_p$ is effectively a free parameter for cross-flow turbine design in confined flow, and should be chosen based on the chord-to-radius ratio to maximize efficiency.

\subsection{Bluff-Body Models for High-Blockage Turbines}

\begin{figure*}[t]
    \centering
    \includegraphics[width = \textwidth]{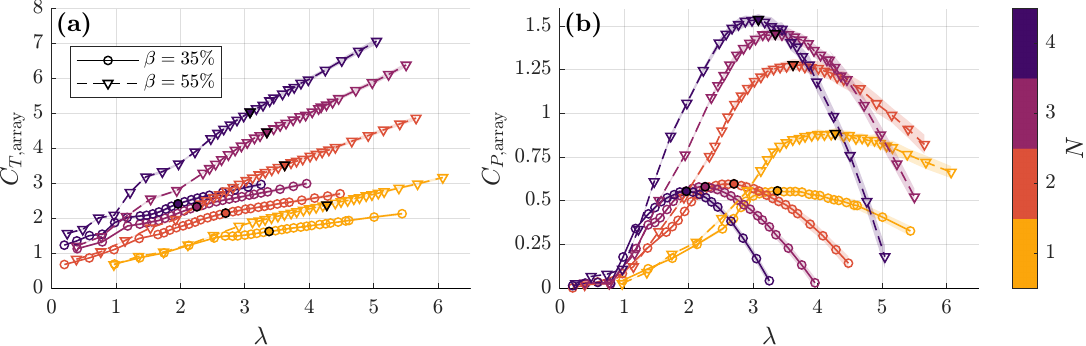}
    \caption{Time-average \textbf{(a)} $C_{T,\mathrm{array}}$ and \textbf{(b)} $C_{P,\mathrm{array}}$ as a function of $N$ and $\lambda$ for $c/R = 0.49$, optimal $\alpha_p$, and $\beta = 35\%$ and $55\%$. The maximum efficiency points are indicated by the filled markers. The shaded regions (which are approximately the same width as the plotted lines at lower $\lambda$) indicate $\pm1$ standard deviation in the cycle-average values.}
    \label{fig:bladeTimeTrends}
\end{figure*}

\begin{figure*}[t]
    \centering
    \includegraphics[width = \textwidth]{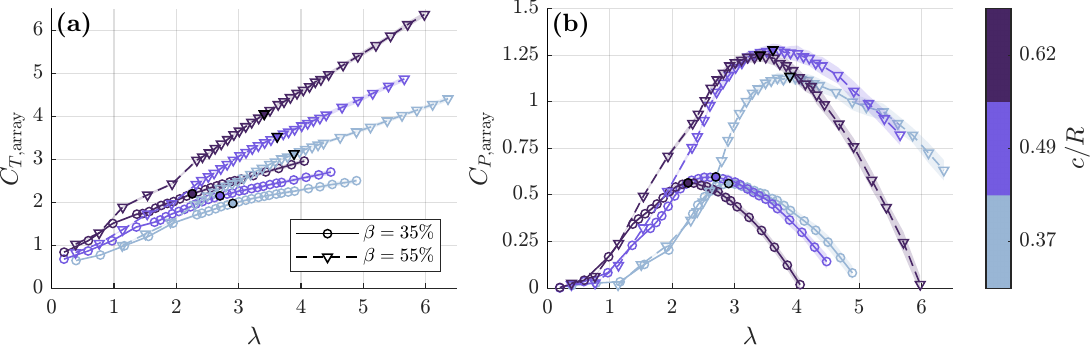}
    \caption{Time-average \textbf{(a)} $C_{T,\mathrm{array}}$ and \textbf{(b)} $C_{P,\mathrm{array}}$ as a function of $c/R$ and $\lambda$ for $N = 2$, optimal $\alpha_p$ at each $c/R$ (\Cref{fig:optPitchLineplot}), and $\beta = 35\%$ and $55\%$. The maximum efficiency points are indicated by the filled markers. The shaded regions (which are approximately the same width as the plotted lines at lower $\lambda$) indicate $\pm1$ standard deviation in the cycle-average values.}
    \label{fig:chordTimeTrends}
\end{figure*}

Thus far, our analysis of trends across the parameter space has focused primarily on array performance at the tip-speed ratio corresponding to maximum efficiency ($\lambda_{\mathrm{opt}}$).
Now, we consider how blockage effects evolve with $\lambda$ across the $\beta$ and geometric parameter space, with a focus on the role of solidity.
At a given $\beta$, an increase in solidity---whether through $N$ or $c/R$---increases the thrust coefficient not only at $\lambda_{\mathrm{opt}}$ (\Cref{fig:ctHeatmap}), but at all $\lambda$.
This trend is shown at $\beta = 35\%$ and $\beta = 55\%$ for a set of representative geometries with various $N$ (at $c/R = 0.49$ and optimal $\alpha_p$) in \Cref{fig:bladeTimeTrends}a and with various $c/R$ (at $N = 2$ and optimal $\alpha_p$) in \Cref{fig:chordTimeTrends}a.
Additionally, an increase in solidity narrows the range of $\lambda$ over which the array produces power, regardless of whether the maximum array efficiency increases or decreases (e.g., \Cref{fig:bladeTimeTrends}b and \Cref{fig:chordTimeTrends}b).
Notably, as solidity increases, confinement affects performance beginning at lower tip-speed ratios.
For example, in \Cref{fig:bladeTimeTrends}b, $C_{P,\mathrm{array}}$ is identical at $35\%$ blockage and $55\%$ blockage up to $\lambda \approx 2.7$ for the single-bladed turbine, but only up to $\lambda \approx 1.4$ for three-bladed turbines, and the ranges of $\lambda$ for which the array thrust coefficient is invariant with $\beta$ are similarly affected.

These results suggest that the confinement effects experienced by the array depend on the ``static'' geometry of its rotors ($\sigma$) and the rotor kinematics ($\lambda$).
Together, $\sigma$ and $\lambda$ influence the resistance to flow through the array, which may be interpreted as the array's resemblance to a bluff body.
Consequently, we consider the combined effects of $\sigma$, $\lambda$, and $\beta$ on array performance using two bluff-body models.
First, the concept of dynamic solidity originally introduced by \citet{araya_transition_2017} is applied to relate rotor geometry to thrust across operating conditions at each $\beta$.
Subsequently, a linear momentum-based \citet{maskell_ec_theory_1963}-inspired bluff-body performance scaling method is applied to relate array thrust to confinement across the geometric and operational space. 
We evaluate the applicability of these models and the connections between them across the parameter space by comparing the thrust characteristics of the turbine array to those of an array of counter-rotating cylindrical shells at a similar range of $\lambda$.

\subsubsection{Dynamic Solidity}
\label{disc:dynamicSolidity}

\citet{araya_transition_2017} defined ``dynamic solidity''  ($\sigma_d$) by relating the advective timescale of the fluid to the timescale of blade passage as
\begin{equation}
    \sigma_d = 1 - \frac{1}{2 \pi \sigma \lambda} \ ,
    \label{eq:dynSolidity}
\end{equation}
For low $\sigma_d$ (i.e., low $\sigma$ or $\lambda$), the gaps between the blades are wide or transit slowly around the rotor circumference, such that the rotor appears relatively porous to the flow and fluid easily passes through the turbine.
As $\sigma_d$ increases, the temporal gaps between the blades narrow, such that it is more difficult for the fluid to pass through, increasing flow resistance by the turbine.
In the limiting case of $\sigma_d \rightarrow 1$, the rotor begins to resemble a solid cylinder.
\citeauthor{araya_transition_2017} showed that increasing the dynamic solidity of a cross-flow turbine at $\beta = 20\%$ resulted in an earlier transition to bluff-body dynamics in the turbine wake.
However, they did not characterize how the thrust coefficients of the turbines developed as a function of dynamic solidity.

This definition of dynamic solidity implicitly chooses the freestream velocity, $U_{\infty}$, as the advective time scale, rather than the actual velocity through the rotors.
While the latter is indeterminate, it is influenced by rotor geometry, rotation rate, and blockage.
Additionally, we note that this model is not interpretable for all turbine geometries and operating states.
As mentioned by \citet{araya_transition_2017}, $\sigma_d$ loses physical meaning when $\lambda < 1/2\pi\sigma$, which results in $\sigma_d < 0$.
Therefore, for the subsequent analysis, operating points with $\sigma_d < 0$ ($\approx 8\%$ of all data points, corresponding to the lower end of the $\lambda$ tested in each experiment) are omitted.
Additionally, even if $\sigma = 1$ (i.e., a solid cylinder), $\sigma_d \neq 1$ for all $\lambda$, even though no fluid can pass through.

\begin{figure*}[t]
    \centering
    \includegraphics[width = \textwidth]{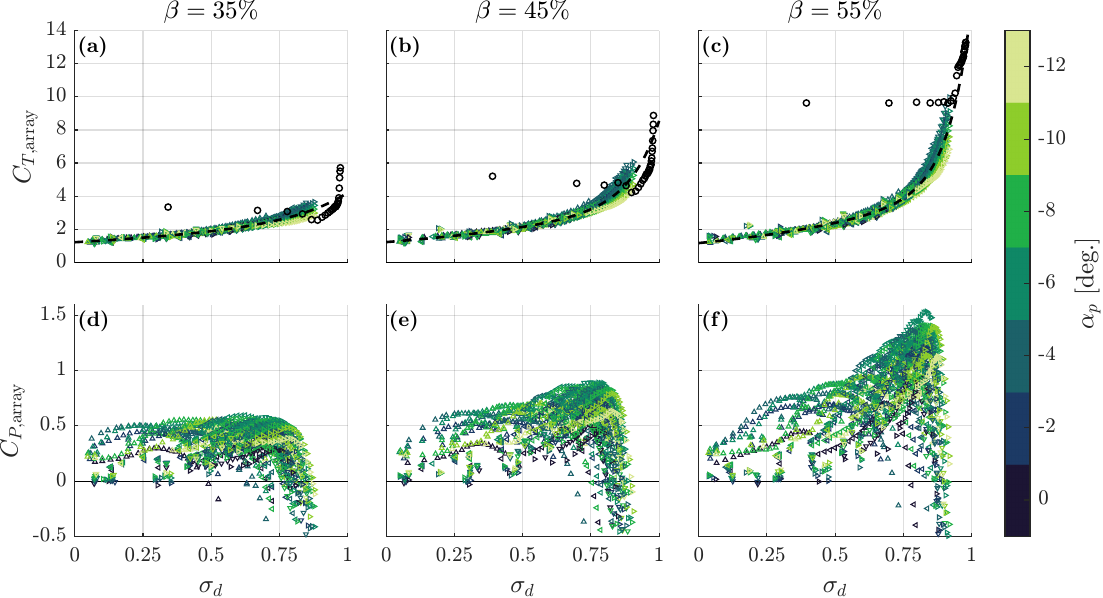}
    \caption{\textbf{(a)--(c)} $C_{T,\mathrm{array}}$ and \textbf{(d)--(f)} $C_{P,\mathrm{array}}$ as a function of dynamic solidity for all geometries and $\beta$.
    Only points with $\sigma_d \geq 0$ are shown.
    The black circles correspond to $C_{T,\mathrm{array}}$ for the cylindrical shells at each $\beta$.
    The dashed lines in (a)--(c) correspond to an exponential fit of $C_{T,\mathrm{array}}$ versus $\sigma_d$ at each $\beta$ (\Cref{eq:solidityFit}; \Cref{tab:solidityFit}).}
    \label{fig:dynamicSolidityTrends}
\end{figure*}

\begin{table}[b]
    \centering
    \setlength{\tabcolsep}{12pt}
    \caption{Coefficient values for \Cref{eq:solidityFit} at each $\beta$.} 
    \label{tab:solidityFit}
    \begin{tabular}{@{}ccccc@{}}
        \toprule
        $\beta$ & $c_1$ & $c_2$ & $c_3$ & $c_4$ \\ \midrule
        35\% & 1.196 & 0.770 & $8.397 \times 10^{-3}$ & 5.263 \\ 
        45\% & 1.213 & 1.086 & $3.814 \times 10^{-4}$  & 9.473 \\ 
        55\% & 1.155 & 1.423 & $5.360 \times 10^{-5}$ & 12.212 \\ \bottomrule 
    \end{tabular}
\end{table}

When the array thrust coefficient is regressed against dynamic solidity, a clear relationship between $\sigma_d$ and $C_{T,\mathrm{array}}$ is observed for all blockage ratios and geometries (\Cref{fig:dynamicSolidityTrends}a-c).
Since arrays at higher blockage ratios produce power over a broader range of $\lambda$, rotors at higher $\beta$ can achieve higher $\sigma_d$ over their operational range, and consequently higher $C_{T,\mathrm{array}}$.
At the highest $\sigma_d$, the measured $C_{T,\mathrm{array}}$ approaches that for the array of cylinders, particularly for $\beta = 55\%$.
The general trend between $C_{T,\mathrm{array}}$ and $\sigma_d$ is well-described at each blockage ratio by a two-term exponential function of the form
\begin{equation}
    C_{T,\mathrm{array}} = c_1 e^{c_2 \sigma_d} + c_3 e^{c_4 \sigma_d} \ ,
    \label{eq:solidityFit}
\end{equation}
\noindent where $c_1 - c_4$ are coefficients selected for best fit at each $\beta$ (\Cref{tab:solidityFit}). The resulting fits are shown as the dashed black lines in \Cref{fig:dynamicSolidityTrends}a--c.

In contrast, there is no clear relationship between the array efficiency and the dynamic solidity (\Cref{fig:dynamicSolidityTrends}d--f).
Although, the highest efficiencies are obtained at higher dynamic solidities as $\beta$ increases, there is considerable spread in $C_{P,\mathrm{array}}$ at each $\sigma_d$.
While this may be exacerbated by the variations in $Re_c$ with $\sigma_d$ (\Cref{disc:ReynoldsEffects}) and differences in blade-level $C_{P,\mathrm{array}}$ at high $\lambda$ due to interactions between the blades and the disk end plates (\Cref{app:diskLoss}), these are likely secondary drivers.
Consequently, as for static solidity (\Cref{fig:solidityTrends}d--f), dynamic solidity is an incomplete descriptor of array efficiency across all operating conditions.

\subsubsection{Maskell-Inspired Bluff-Body Scaling}
\label{disc:bluffBodyScaling}

\begin{figure*}[t]
    \centering
    \includegraphics[width = \textwidth]{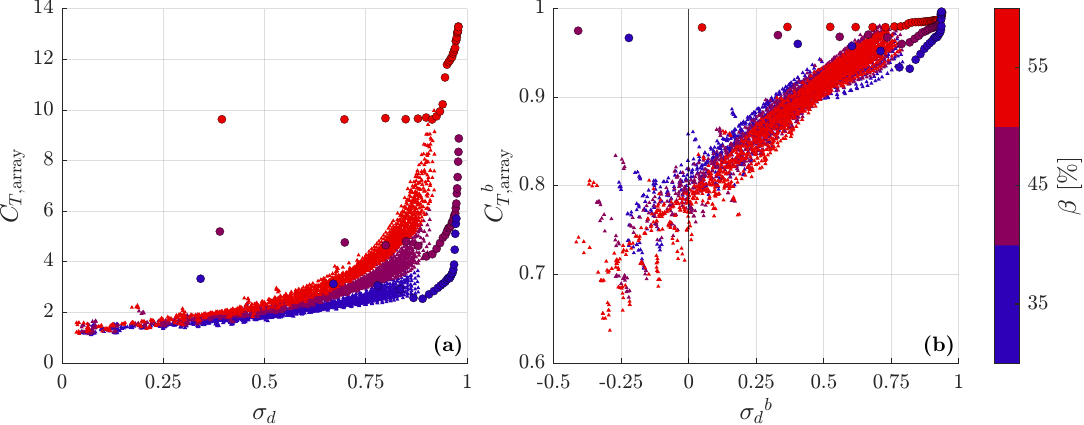}
    \caption{\textbf{(a)} $C_{T,\mathrm{array}}$ as a function of dynamic solidity for all geometries at all $\beta$. \textbf{(b)} Bluff-body scaled thrust vs bluff-body scaled dynamic solidity for all geometries at all $\beta$. Only points with $\sigma_d \geq 0$ are shown. The larger circles in each panel correspond to the thrust measurements for the cylindrical shells at each $\beta$.}
    \label{fig:bluffBodyScaling}
\end{figure*}

Even at constant $\sigma_d$, $C_{T,\mathrm{array}}$ increases with $\beta$ (\Cref{fig:bluffBodyScaling}a), indicating that dynamic solidity alone is not fully descriptive of the flow dynamics.
Consequently, to interpret the relationship between the array thrust and the blockage ratio across the geometric and operational space, we turn to a linear momentum and bluff-body model for cross-flow turbine dynamics that was previously applied to this array by \citet{hunt_experimental_2024a}.
\citeauthor{hunt_experimental_2024a} considered a cross-flow turbine array with a single rotor geometry ($N=2$, $c/R=0.49$, $\alpha_p = -6^{\circ}$, and blade-end strut supports) and characterized the performance and near-wake flow-field across a similar range of blockage ratios ($\beta = 30\% - 55\%$).
The authors then modeled the array using the open-channel linear momentum actuator disk theory (LMADT) method developed by \citet{houlsby_application_2008}, which treats the array as an porous, non-rotating disk with the same $\beta$ as the array that extracts momentum from a uniform inflow with velocity $U_{\infty}$ and depth $h$.
A detailed derivation of the open-channel LMADT model is provided by \citet{houlsby_application_2008} and \citet{houlsby_power_2017}.
A complete description of the adaption of this model to the experimental array in this study is provided by \citet{hunt_experimental_2024a}.
Using the measured $C_{T,\mathrm{array}}$ and flow characteristics as inputs, the LMADT model was used to estimate the wake velocity ($u_w$) and the velocity of the fluid that bypasses the array ($u_b$) by numerically solving
\begin{equation}
    u_w = \frac{Fr_h^2 u_b^4 - (4 + 2Fr_h^2) U_{\infty}^2 u_b^2 + 8 U_{\infty}^3 u_b - 4U_{\infty}^4 + 4\beta C_{T,\mathrm{array}} U_{\infty}^4 + Fr_h^2 U_{\infty}^4}
               {-4Fr_h^2 u_b^3 + (4Fr_h^2 + 8) U_{\infty}^2 u_b - 8U_{\infty}^3} \  
    \label{eq:uwBlockage}
\end{equation}
 and 
\begin{equation}
    u_w = \sqrt{u_b^2 - C_{T,\mathrm{array}}U_{\infty}^2} \ 
    \label{eq:uwThrust}
\end{equation}
\noindent for $u_w$ and $u_b$.
\citeauthor{hunt_experimental_2024a} found that $u_b$ was an excellent predictor of the velocity in the near-wake bypass region, as compared to measurements with particle image velocimetry across various $\beta$ and $\lambda$.
Subsequently, following \citet{whelan_freesurface_2009}'s adaptation of \citet{maskell_ec_theory_1963}'s bluff-body theory, \citeauthor{hunt_experimental_2024a} scaled the array-average power and thrust coefficients by this bypass velocity as
\begin{equation}
    C_{P,\mathrm{array}}^{\ \: b} = C_{P,\mathrm{array}} {\left( \frac{U_{\infty}}{u_b} \right)}^3 ,
    \label{eq:cpBluff}
\end{equation}
\begin{equation}
    C_{T,\mathrm{array}}^{\ \: b} = C_{T,\mathrm{array}} {\left( \frac{U_{\infty}}{u_b} \right)}^2 ,
    \label{eq:ctBluff}
\end{equation}
\begin{equation}
    {\lambda}^b = \lambda\left( \frac{U_{\infty}}{u_b} \right) ,
    \label{eq:TSRBluff}
\end{equation}
\noindent The resulting $C_{P,\mathrm{array}}^{\ \: b}-\lambda^b$ and $C_{T,\mathrm{array}}^{\ \: b}-\lambda^b$ curves were self-similar across the tested $\beta$ and $\lambda$, indicating that the array power and thrust scale with the bypass velocity at high blockage.

Given the 60 unique combinations of $c/R$, $\alpha_p$, and $N$ tested at each $\beta$ in this study, we can test if this self-similarity in array performance generalizes to a broader geometric space.
Here, we focus on the measured $C_{T,\mathrm{array}}$ and corresponding bypass-scaled thrust ($C_{T,\mathrm{array}}^{\ \: b}$) for each geometric configuration, as the thrust coefficient is the only performance metric that influences the value of $u_b$, and consequently, the effectiveness of \citet{maskell_ec_theory_1963}-inspired bluff-body performance scaling for a particular geometry.
For each rotor geometry, $u_b$ is calculated from the measured $C_{T,\mathrm{array}}$, $U_{\infty}$, and $Fr_h$ at each $\beta$ and $\lambda$ using \Cref{eq:uwBlockage,eq:uwThrust}.
Then, $C_{T,\mathrm{array}}^{\ \: b}$ and ${\lambda}^b$ are calculated from \Cref{eq:ctBluff,eq:TSRBluff}, respectively, for that rotor geometry and $\beta$.
Finally, to incorporate the influence of rotor geometry on the array thrust, at each $\beta$ we calculate a bypass-scaled dynamic solidity as
\begin{equation}
    {\sigma_d}^b = 1 - \frac{1}{2 \pi \sigma \lambda^b}
    \label{eq:dynSolidityBluff}
\end{equation}
\noindent As for $\sigma_d$, ${\sigma_d}^b < 0$ when $\lambda_b < 1/2\pi\sigma$, which can occur even if the corresponding $\sigma_d > 0$.

The $C_{T,\mathrm{array}}^{\ \: b}$ for all geometries and $\beta$ are plotted against the corresponding ${\sigma_d}^b$ in \Cref{fig:bluffBodyScaling}b, and show that Maskell-inspired bluff-body scaling results in self-similar $C_{T,\mathrm{array}}-\sigma_d$ for all geometries and operating conditions across the tested blockage ratios.
The greatest scatter in $C_{T,\mathrm{array}}^{\ \: b}$ is observed at lower ${\sigma_d}^b$, where resistance to flow through the array is relatively low and $C_{T,\mathrm{array}}$ varies weakly with both $\beta$ and $\sigma_d$ (\Cref{fig:bluffBodyScaling}a).
However, as ${\sigma_d}^b$ increases and the array thrust approaches that of a bluff body, the collapse in $C_{T,\mathrm{array}}^{\ \: b}$ generally improves, with residual scatter attributed to the unique influences of $c/R$, $N$, and $\alpha_p$.
Additionally, the self-similar relationship between $C_{T,\mathrm{array}}^{\ \: b}$ and ${\sigma_d}^b$ is linear across the parameter space, with $C_{T,\mathrm{array}}^{\ \: b}$ trending toward unity as ${\sigma_d}^b$ increases.
The limiting case of $C_{T,\mathrm{array}}^{\ \: b} = 1$ corresponds to $C_{T,\mathrm{array}} = (u_b/U_{\infty})^2$ (\Cref{eq:ctBluff}), which results in $u_w = 0$ from \Cref{eq:uwThrust}.
This result is aligned with the original bluff-body theory of \citet{maskell_ec_theory_1963}, who suggested that the thrust on a bluff-body is directly related to the acceleration of the bypass flow, and assumed zero velocity in the wake in their derivation. The alignment between the experimental results and theory is further supported by the agreement between the turbine $C_{T,\mathrm{array}}^{\ \: b}$ and cylinder $C_{T,\mathrm{array}}^{\ \: b}$ (i.e., a true bluff body) at higher ${\sigma_d}^b$ (\Cref{fig:bluffBodyScaling}b).
A discussion of how these results inform self-similarity in array thrust across $\beta$ for specific geometric configurations is provided as supplementary material.

The result in \Cref{fig:bluffBodyScaling}b suggests a remarkable compatibility between these two different bluff-body models for cross-flow turbine dynamics across a variety of rotor geometries, operating conditions, and $\beta$.
\citet{maskell_ec_theory_1963}-inspired bluff-body scaling uses open-channel linear momentum theory to model the channel dynamics as a function of $\beta$ and the thrust on the array, but does not explicitly consider the effects of turbine geometry or rotation since the array is treated as an actuator disk.
Conversely, \citet{araya_transition_2017}'s dynamic solidity model relates the rotors' geometry and kinematics to flow resistance via a scaling argument, but does not formally consider the blockage ratio, turbine-channel interactions, or the forcing on the turbine.
Consequently, it is notable that, while neither model provides a complete description, the combination is fully descriptive of how $C_{T,\mathrm{array}}$ evolves with rotor geometry, rotor kinematics, and the blockage ratio.

\section{Conclusions}
\label{sec:conclusion}

The objective of this work is to experimentally investigate how the optimal rotor geometry for a cross-flow turbine depends on confinement, with a focus on the upper end of practically achievable blockage ratios.
Three geometric parameters are considered in this study: the number of blades ($N = 1-4$), the chord-to-radius ratio ($c/R = 0.37 - 0.62$), and the preset pitch angle ($\alpha_p = -12^{\circ} - 0^{\circ}$).
Using an array of two counter-rotating cross-flow turbines, the power and force characteristics of 60 unique rotor geometries are evaluated at $\beta = 35\% - 55\%$ in a recirculating water tunnel.
While the maximum efficiency and corresponding thrust coefficient and tip-speed ratio all increase with the $\beta$ regardless of rotor geometry, the effects of blockage manifest differently depending on the geometric configuration of the rotor.
Most notably, as the number of blades increases, the power and thrust coefficients increase with $\beta$ at a faster rate and higher maximum efficiencies are achieved, whereas an inverse relationship between maximum efficiency and blade count is typically observed at lower $\beta$ \citep{li_effect_2015, araya_transition_2017, miller_solidity_2021, hunt_experimental_2024}.
In contrast, neither the optimal chord-to-radius ratio nor the optimal preset pitch angle depend on $\beta$ in the range tested and the optimal preset-pitch angle primarily depends on the chord-to-radius ratio.

At a given $\beta$, the array thrust coefficient at maximum efficiency is shown to be primarily a function of solidity, a geometric parameter that combines $N$ and $c/R$.
This relationship between $C_{T,\mathrm{array}}$ and rotor geometry is extended to all operating conditions through the concept of dynamic solidity ($\sigma_d$), a parameter introduced by \citet{araya_transition_2017} that combines the ``static'' solidity of the rotor and the tip-speed ratio to describe a turbine's resemblance to a bluff-body.
The array thrust coefficient is shown to increase exponentially with dynamic solidity, and this effect is augmented as the blockage ratio increases.
To relate the array thrust and the blockage ratio, a Maskell-inspired linear momentum and bluff-body model is applied, in which the bypass velocity around the array in each experiment is predicted from the measured thrust and flow conditions.
When $C_{T,\mathrm{array}}$ and $\sigma_d$ are normalized by this bypass velocity, the relationship between these parameters is linearly self-similar across the tested geometries, tip-speed ratios, and blockage ratios.
This result highlights a remarkable compatibility between these two distinct bluff-body models---one which focuses solely on the dynamics of the array-channel interaction, and another that solely focuses on the rotor geometry and kinematics---and demonstrates their suitability for describing how rotor geometry influences the blockage effects experienced by the array.
In contrast, the relationship between $C_{P,\mathrm{array}}$ and solidity is weaker, though since the blockage effects experienced by the array increase with the thrust coefficient (and thus, dynamic solidity), higher maximum efficiencies are generally obtained with higher $\sigma_d$ as $\beta$ increases.
The specific geometric configuration of the rotor influences $C_{P,\mathrm{array}}$ at a given $\sigma_d$, indicating that solidity alone does not fully describe array power in high-blockage flows.

Since blockage effects are a function of both the channel blockage ratio and the thrust coefficient, the design process for high-blockage turbines is expected to be driven by the optimal solidity that most effectively exploits confinement at a given $\beta$.
Increasing the number of blades is the simplest and most favorable pathway for achieving the optimal solidity, as it increases the power generated while simultaneously reducing the amplitude of cyclic loading.
The optimal solidity may also be achieved by changing the chord-to-radius ratio, although the chosen $c/R$ will likely be influenced structural limitations on the rotor and blade size, as well as the target Reynolds number.
In contrast, the preset pitch angle is a free parameter in the design process, and should be chosen based on the chord-to-radius ratio to maximize efficiency.
We note that, due to the effects of blockage on the local Reynolds number, as well as the influence of the Reynolds number on the optimal $c/R$ \citep{bianchini_design_2015, hunt_experimental_2024}, the trends observed in this study may change as Reynolds independence is approached.
Additionally, several other geometric parameters, such as the blade profile, shape, and surface roughness can also influence performance and are worthy of future investigation.
Nonetheless this study provides the first experimental examination of the influence of blockage on optimal rotor geometry and informs design principles for cross-flow turbines in high-blockage flows.


\begin{acknowledgments}
    This work was supported by the United States Advanced Research Projects Agency – Energy (ARPA-E) under award number DE-AR0001441.
    The authors would like to thank the Alice C. Tyler Charitable Trust and the United States Department of Defense Naval Facilities Engineering Command (NAVFAC) for supporting improvements to the experimental facilities and turbine instrumentation at the University of Washington.
\end{acknowledgments}

\section*{Data Availability Statement}
    The data that support the findings of this study are openly available in the Dryad data repository at \url{https://doi.org/10.5061/dryad.1c59zw45d}\citep{hunt_data_2024b}.
    The MATLAB code used to implement open-channel linear momentum theory is available on GitHub at \url{https://github.com/aidan-hunt/turbine-confinement-models}.

\section*{Supplementary Material}
The following are provided as supplementary material:
velocity profiles in the flume with the array absent;
tables of the range of measured $\beta$, $Re_D$, and $Fr_h$ in each experiment;
a brief discussion of uncertainty analysis;
contours of the rotor-averaged lateral force coefficient at $\lambda_{\mathrm{opt}}$; $C_{T,\mathrm{array}}-\lambda$ curves for the cylindrical shells; $C_{T,\mathrm{array}}^{\ \: b}-\lambda^b$ curves for a set of representative geometries.
    
\appendix
\numberwithin{equation}{section}
\numberwithin{figure}{section}
\numberwithin{table}{section}

\newpage
\section{Variation in Parameters with Geometric Configuration}
\label{app:radVar}

\begin{table}[h]
    \setlength{\tabcolsep}{6pt}
    \centering
    \caption{Quarter-chord radius $R$, outermost radius $R'$, and derived quantities for each combination of chord length and preset pitch. The listed $\beta$ and $Re_D$ correspond to variations in the \textit{nominal} blockage ratio and Reynolds number as a function of $c/R$ and $\alpha_p$.}
    \label{tab:radVar}
    \resizebox{\textwidth}{!}{
        \begin{tabular}{@{}ccccccccc@{}}
            \toprule
            $c$ [cm] & $\alpha_p$ [deg.] & $R$ [cm] & $R'$ [cm] & $R'/R$ & $c/R$ & $A_{\mathrm{turbines}}$ [cm$^2$] & Nominal $\beta$ [\%] & Nominal $Re_D$ [$\times10^-5$] \\ \midrule
            \multirow{5}{*}{5.565} & 0 & 14.73 & 15.31 & 1.04 & 0.378 & 1317.0 & 34.0, 43.8, 53.5 & 1.59 \\
             & -2 & 14.93 & 15.43 & 1.03 & 0.373 & 1326.6 & 34.3, 44.1, 53.9 & 1.61 \\
             & -4 & 15.08 & 15.58 & 1.03 & 0.369 & 1340.1 & 34.6, 44.5, 54.4 & 1.63 \\
             & -6 & 15.07 & 15.58 & 1.03 & 0.369 & 1339.6 & 34.6, 44.5, 54.4 & 1.62 \\
             & -8 & 15.04 & 15.57 & 1.04 & 0.370 & 1338.9 & 34.6, 44.5, 54.4 & 1.62 \\ \midrule
            \multirow{5}{*}{7.420} & -2 & 14.93 & \multirow{5}{*}{15.75} & 1.05 & 0.497 & \multirow{5}{*}{1354.5} & \multirow{5}{*}{35.0, 45.0, 55.0} & 1.61 \\
             & -4 & 15.08 &  & 1.04 & 0.492 &  &  & 1.63 \\
             & -6 & 15.07 &  & 1.05 & 0.492 &  &  & 1.62 \\
             & -8 & 15.04 &  & 1.05 & 0.493 &  &  & 1.62 \\
             & -10 & 15.01 &  & 1.05 & 0.494 &  &  & 1.62 \\ \midrule
            \multirow{5}{*}{9.275} & -4 & 15.08 & 16.16 & 1.07 & 0.615 & 1390.1 & 35.9, 46.2, 56.4 & 1.63 \\
             & -6 & 15.07 & 15.93 & 1.06 & 0.616 & 1369.6 & 35.4, 45.5, 55.6 & 1.62 \\
             & -8 & 15.04 & 15.93 & 1.06 & 0.617 & 1370.3 & 35.4, 45.5, 55.6 & 1.62 \\
             & -10 & 15.01 & 15.94 & 1.06 & 0.618 & 1371.2 & 35.4, 45.6, 55.7 & 1.62 \\
             & -12 & 14.98 & 15.95 & 1.07 & 0.619 & 1372.1 & 35.5, 45.6, 55.7 & 1.61 \\ \bottomrule
        \end{tabular}
    }
\end{table}

\newpage
\section{Limitations of Blade-Level Performance Estimation at High Blockage}
\label{app:diskLoss}

\begin{figure*}[t]
    \centering
    \includegraphics[width = \textwidth]{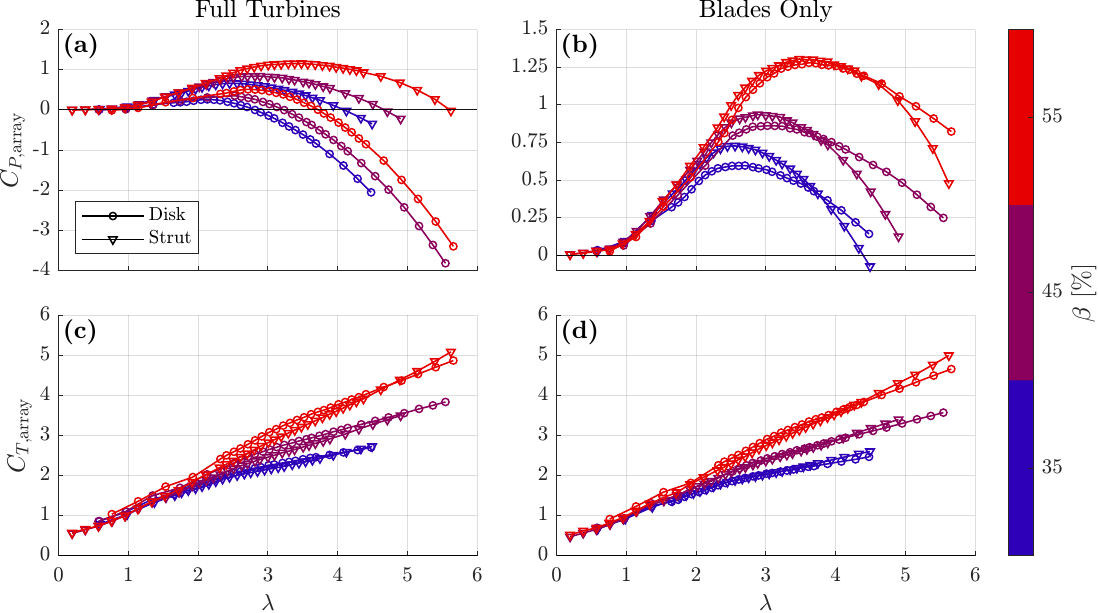}
    \caption{\textbf{(a)} $C_{P,\mathrm{array}}$ calculated using the full turbine measurements (i.e., including support structure losses) compared to \textbf{(b)} $C_{P,\mathrm{array}}$ accounting for support structure losses using \Cref{eq:cpBlade} for a single turbine geometry ($N=2$, $c/R = 0.49$, $\alpha_p = -6^{\circ}$) across the tested $\beta$.
    The performance curves compare turbines with disk end plates as in this study (\Cref{sec:methods}) to turbines with blade-end strut supports in prior work \citep{hunt_experimental_2024a}). \textbf{(c)} $C_{T,\mathrm{array}}$ for the full turbines compared to \textbf{(d)} $C_{T,\mathrm{array}}$ if the thrust on the support structures is subtracted using an analogous equation to \Cref{eq:cpBlade}.}
    \label{fig:diskVsStrut}
\end{figure*}

As described in \Cref{sec:methods}, disk end plates are used to connect blades to the central driveshaft of each rotor, which enables cost-effective testing of numerous rotor geometries with common sets of blade support structures.
However, these end plates generate considerable parasitic torque, which significantly reduces the measured array-average efficiency relative to the same rotor geometry with blades supported by thin, hydrodynamic struts, as shown for a representative geometry at various $\beta$ in \Cref{fig:diskVsStrut}a.
The latter choice of support structure more closely approximates the performance of the blades alone, but would necessitate a unique strut assembly to be fabricated for each geometric configuration, each of which would incur a relatively high cost.
In contrast, the measured array thrust coefficient does not depend significantly on whether struts or disk end plates are used to support the blades (\Cref{fig:diskVsStrut}c), implying that the thrust measured for the full turbines is dominated by the force on the blades.

To account for the effect of support structure and generalize the results, the efficiency of the blades only (\Cref{fig:diskVsStrut}b) is estimated using a linear superposition method (\Cref{eq:cpBlade}), in which the $C_P$ measured for the support structures only (i.e., an array of bladeless turbines) is subtracted from the full turbine $C_P$ at the same tip-speed ratio.
This technique, which was shown by \citet{strom_impact_2018} to yield similar blade-only $C_P$ for turbines with different types of support structures at low blockage ($\beta = 11\%$), assumes secondary interactions between the blades and the support structures are negligible.
In other words, although the choice of support structure changes the net power produced by the turbine due to the associated parasitic torque, this method assumes that the power produced by the blades themselves is unaffected by the presence of the support structures.
While blade-only $C_{P,\mathrm{array}}$ across the parameter space is presented in \Cref{sec:results,sec:discussion}, support structure subtraction is not employed for $C_{T,\mathrm{array}}$, although the results of an analogous method to \Cref{eq:cpBlade} for thrust are shown in \Cref{fig:diskVsStrut}d.

\begin{figure*}[t]
    \centering
    \begin{minipage}[t]{0.475\textwidth}
        \centering
        \includegraphics[width = \textwidth]{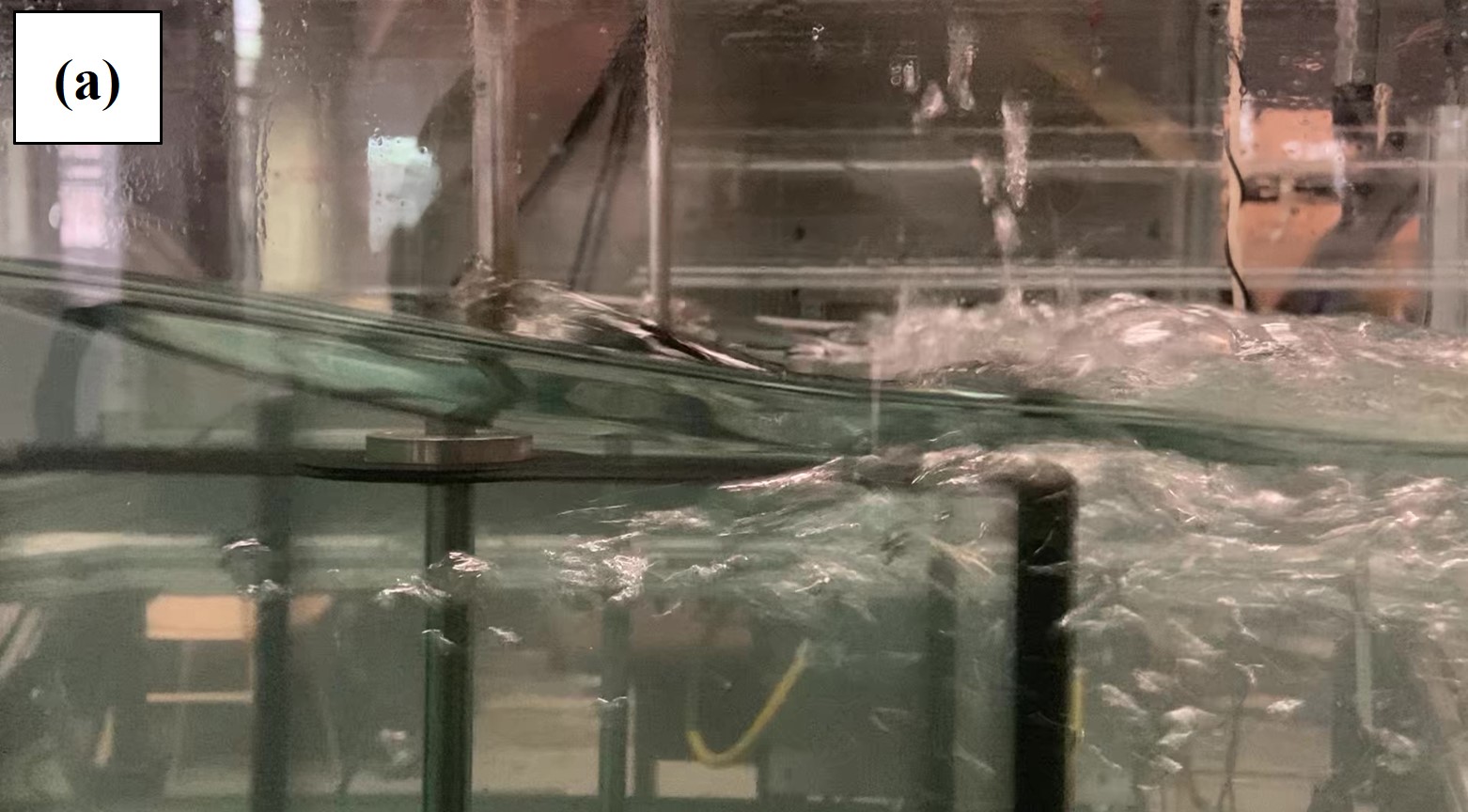}
    \end{minipage} \hfill
    \begin{minipage}[t]{0.475\textwidth}
        \centering
        \includegraphics[width = \textwidth]{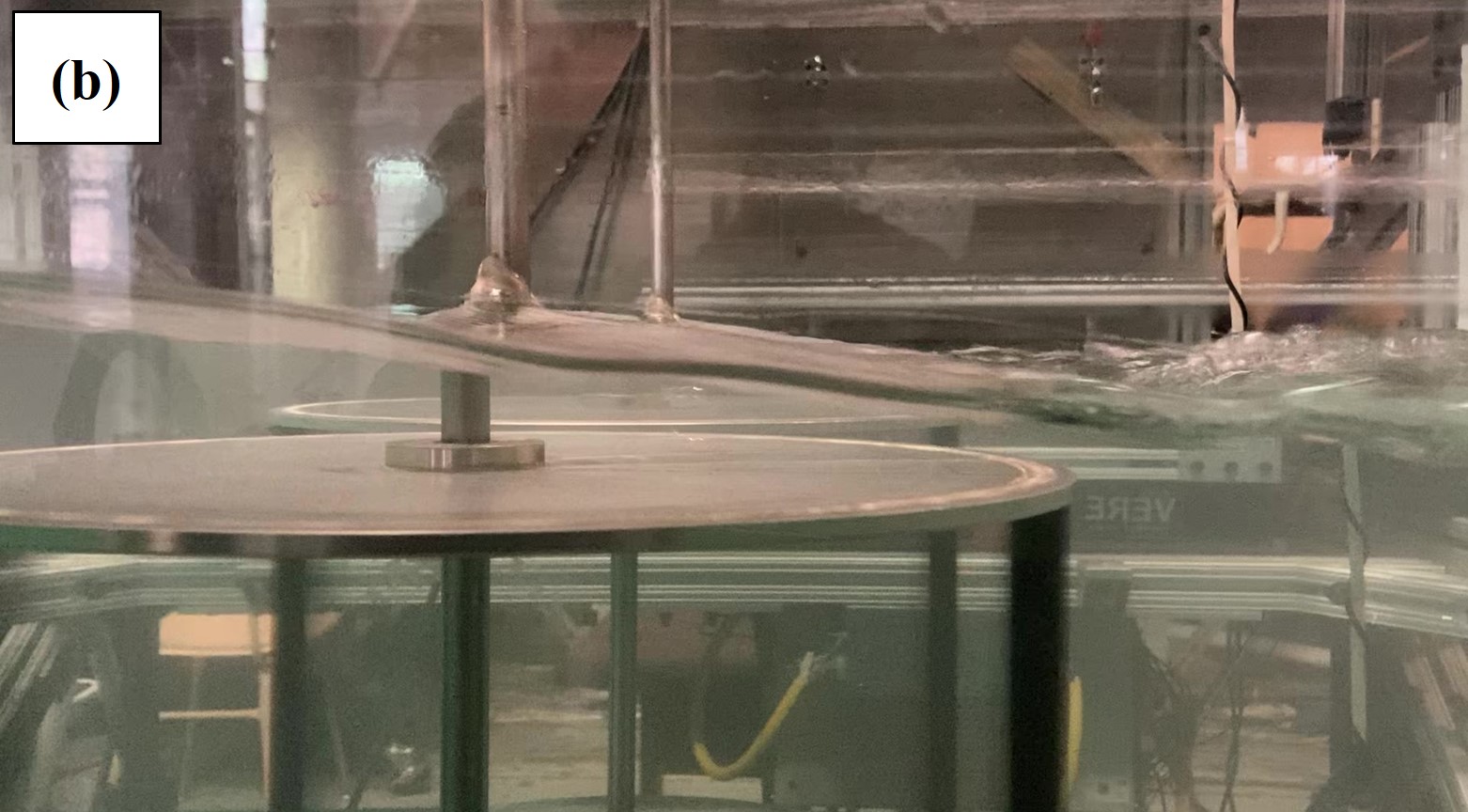}
    \end{minipage}
    \caption{Interaction between a turbine ($N=2$, $c/R = 0.49$, $\alpha_p = -6^{\circ}$) and free surface at $\beta = 55\%$ and $\lambda \approx 5.6$, compared between \textbf{(a)} turbines with blade-end struts \citep{hunt_experimental_2024a} and \textbf{(b)} turbines with disk end plates. Flow in the images moves from left to right.}
    \label{fig:ventilation}
\end{figure*}

Estimating blade-level efficiency in this way yields moderate agreement in blade-only $C_{P,\mathrm{array}}$ between arrays with struts or disk end plates (\Cref{fig:diskVsStrut}b), and the blade-only performance curves have greater dependency on the type of support structure used than observed in prior work \citep{strom_impact_2018}.
For example, at most $\lambda$, blade-only $C_{P,\mathrm{array}}$ is lower for arrays with disk supports than for strut supports, except for at higher $\lambda$ beyond the performance peak where blade-only $C_{P,\mathrm{array}}$ is elevated for turbines with disk end plates.
Therefore, because \Cref{eq:cpBlade} not does yield equivalent blade-only $C_{P,\mathrm{array}}-\lambda$ curves between the disk turbines and the strut turbines at these $\beta$, the effects of support structures on the blades are not completely represented by the $C_{P}$ measured for the support structures alone.
We hypothesize that this is because, at high blockage, the blade thrust significantly alters the flow field around the turbine and changes the nature of the hydrodynamic interactions between the blades and supports, such that the choice of support structure influences the power produced by the blades.
At low-to-moderate $\lambda$ at a given $\beta$, the lower blade-only $C_{P,\mathrm{array}}$ for disk turbines implies that such interactions are detrimental to performance.
In contrast, at the upper end of the tested $\lambda$ at each $\beta$, the higher blade-only $C_{P,\mathrm{array}}$ of disk turbines may imply a favorable hydrodynamic interaction between the blades and the disk end plates.
One such hydrodynamic benefit of the disk end plates is apparent in the interaction between the array and the free surface at the high-blockage, high-rotation-rate case of $\beta = 55\%$ and $\lambda \approx 5.6$ shown in \Cref{fig:ventilation}.
For turbines with blade-end struts under these conditions (\Cref{fig:ventilation}a), the free surface is pulled down into the rotor and significant ventilation occurs, which degrades turbine performance \citep{hunt_experimental_2024a}.
In contrast, the disk end plates suppress ventilation at this $\lambda$ and $\beta$ (\Cref{fig:ventilation}b), such that no air is entrained in the rotor.

While \Cref{fig:ventilation} highlights an extreme case where the differing effects of the disks and the struts on turbine hydrodynamics are plainly visible, it is likely that more subtle interactions between the freestream flow, the blades, and the support structures influence blade performance at lower $\beta$ and $\lambda$ as well.
Furthermore, we hypothesize that the increase in the apparent agreement between blade-only $C_{P,\mathrm{array}}$ for strut turbines and the disk turbines with increasing $\beta$ (\Cref{fig:diskVsStrut}b) is driven by the development of beneficial interactions between the blades and disk end plates, rather than improved accuracy of the support structure subtraction model.
This is not to say that turbines with disk end plates become more preferable than those with struts at higher blockage; we emphasize that, at all $\beta$, the net power produced by the array of disk turbines is far lower than that produced by the array of strut turbines (\Cref{fig:diskVsStrut}a).
Rather, under these conditions, the choice of support structure influences the power produced by the blades in a manner that evolves with $\beta$ and $\lambda$, such that the choice of support structure changes the estimate of blade-only $C_{P,\mathrm{array}}$ obtained using \Cref{eq:cpBlade}.
As discussed in \Cref{disc:bluffBodyScaling}, this dependency may contribute to some of the scatter in $C_{P,\mathrm{array}}$ as a function of $\sigma_d$.
Further, when using these results to validate numerical and reduced-order models, it is important to be mindful of how the choice of support structure affects the estimated values for blade-only $C_{P,\mathrm{array}}$.
Nonetheless, since disk end plates are used for all turbines in this study, we do not expect the choice of support structure to significantly affect the trends observed in maximum blade-only $C_{P,\mathrm{array}}$.

\newpage
\bibliography{bgeom_refs}

\FloatBarrier
\newpage
\nolinenumbers
\pagestyle{plain}

\clearpage
\pagenumbering{arabic}
\renewcommand*{\thepage}{S\arabic{page}}

\renewcommand{\thefigure}{S.\arabic{figure}}
\setcounter{figure}{0}
\renewcommand{\thetable}{S.\arabic{table}}
\setcounter{table}{0}
\renewcommand{\theequation}{S.\arabic{equation}}
\setcounter{equation}{0}

\include{supp_title}

\include{supp_velProfiles}

\include{supp_flowVar}

\include{supp_uncertainty}

\include{supp_lateral}

\include{supp_cylinders}

\include{supp_scaling}

\end{document}

%% file: supp_title.tex
\begin{centering}
    \Large
    \textbf{Supplemental information for ``Performance characteristics and bluff-body modeling of high-blockage cross-flow turbine arrays with varying rotor geometry''}

    \vspace{1cm}

    \large
    Aidan Hunt\textsuperscript{1,$\ast$}, Gregory Talpey\textsuperscript{1}, Gemma Calandra\textsuperscript{1}, and Brian Polagye\textsuperscript{1}

    \normalsize
    \textsuperscript{1} Department of Mechanical Engineering, University of Washington.
    
    \textsuperscript{$\ast$} To whom correspondence should be addressed: ahunt94@uw.edu.

    \vspace{2cm}

\end{centering}

This PDF file contains the following supplementary information:

\begin{itemize}[align=right,
                  leftmargin=2em,
                  topsep=0.1em,
                  itemsep=0.1em]
    \item velocity profiles in the flume with the array absent,
    \item tables of the range of measured $\beta$, $Re_D$, and $Fr_h$ in each experiment,
    \item a brief discussion of uncertainty analysis,
    \item contours of the rotor-averaged lateral force coefficient at $\lambda_{\mathrm{opt}}$,
    \item $C_{T,\mathrm{array}}-\lambda$ curves for the cylindrical shells,
    \item $C_{T,\mathrm{array}}^{\ \: b}-\lambda^b$ curves for a set of representative geometries.
\end{itemize}

%% file: supp_velProfiles.tex
\section*{Velocity Profiles}

   \begin{figure}[hbt!]
        \centering
        \includegraphics[width=\textwidth]{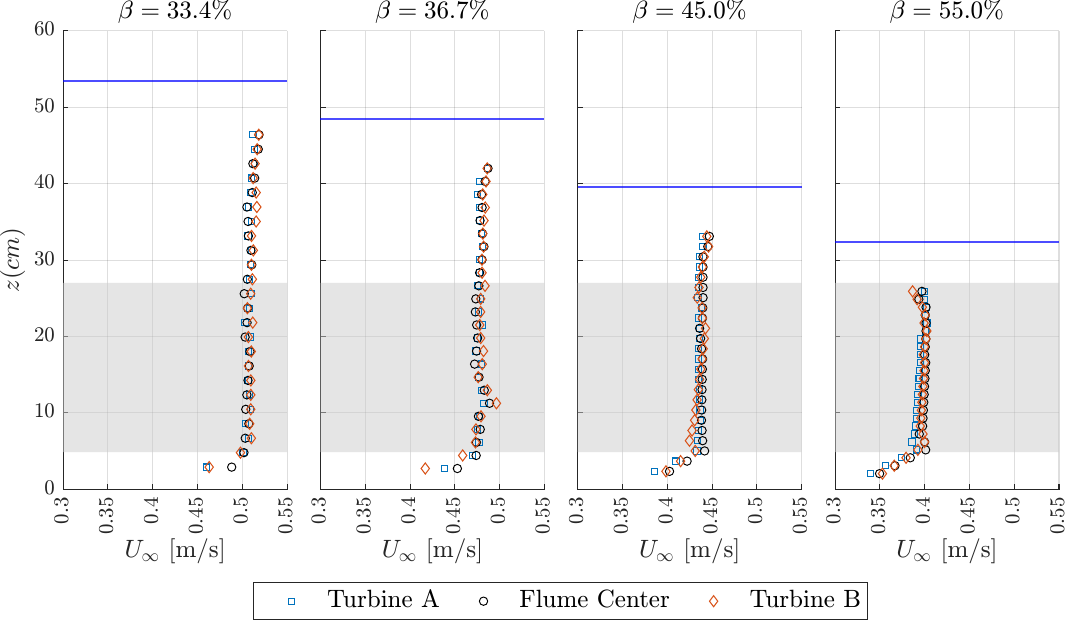}
        \caption{Vertical velocity profile measured in the flume without the array present at target flow conditions corresponding to various blockage ratios at $Re_D = 1.62\!\times\!10^5$ and $Fr_h = 0.219$. Profiles were collected at streamwise positions located five diameters upstream of the array, laterally located at the flume centerline as well as directly upstream of each turbine's axis of rotation (i.e., $0$, $-w/4$, and $w/4$ in Figure 4). The gray shaded region denotes the array's position in the water column during performance experiments. The dark blue line in each tile indicates the nominal dynamic depth in the flume corresponding to each $\beta$. While velocity profiles were not collected at $\beta = 35\%$, similar trends are expected based on the profiles at $\beta=33.4\%$ and $\beta = 36.7\%$.}
        \label{fig:velProfiles}
    \end{figure}

%% file: supp_flowVar.tex
\newpage

\begin{sidewaystable}[h]
    \setlength{\tabcolsep}{6pt}
    \centering
    \caption{Range of the time-averaged values of $\beta$, $Fr_h$, and $Re_D$  (averaged at each $\lambda$) measured for each geometric configuration tested at $\beta = 35\%$.}
    \label{tab:flowVar35}
    \begin{adjustbox}{totalheight=0.5\textheight, center}
    \begin{tabular}{@{}|cc|cccc|cccc|cccc|@{}}
    \toprule
     &  & \multicolumn{4}{c|}{Measured $\beta$ [\%]} & \multicolumn{4}{c|}{Measured $Fr_h$} & \multicolumn{4}{c|}{Measured $Re_D$ [$\times10^-5$]} \\ \midrule
    $c/R$ & $\alpha_p$ [deg.] & $N=1$ & $N=2$ & $N=3$ & $N=4$ & $N=1$ & $N=2$ & $N=3$ & $N=4$ & $N=1$ & $N=2$ & $N=3$ & $N=4$ \\ \midrule
    \multirow{5}{*}{0.37} & 0 & 34.1 - 34.2 & 34.1 - 34.2 & 34.0 - 34.2 & 34.0 - 34.2 & 0.213 - 0.217 & 0.214 - 0.216 & 0.214 - 0.218 & 0.211 - 0.217 & 1.55 - 1.58 & 1.54 - 1.56 & 1.55 - 1.58 & 1.53 - 1.58 \\
     & -2 & 34.3 - 34.5 & 34.3 - 34.4 & 34.2 - 34.3 & 33.9 - 34.1 & 0.213 - 0.217 & 0.214 - 0.216 & 0.214 - 0.218 & 0.212 - 0.215 & 1.57 - 1.60 & 1.57 - 1.58 & 1.57 - 1.59 & 1.55 - 1.57 \\
     & -4 & 34.6 - 34.8 & 34.6 - 34.8 & 34.6 - 34.7 & 33.8 - 34.0 & 0.215 - 0.219 & 0.214 - 0.218 & 0.215 - 0.218 & 0.212 - 0.217 & 1.60 - 1.62 & 1.59 - 1.61 & 1.60 - 1.62 & 1.54 - 1.57 \\
     & -6 & 34.7 - 34.9 & 34.7 - 34.8 & 34.6 - 34.7 & 33.8 - 33.9 & 0.217 - 0.219 & 0.215 - 0.218 & 0.215 - 0.218 & 0.211 - 0.216 & 1.62 - 1.63 & 1.60 - 1.61 & 1.59 - 1.61 & 1.53 - 1.56 \\
     & -8 & 34.7 - 34.9 & 34.6 - 34.8 & 34.6 - 34.7 & 33.7 - 33.8 & 0.216 - 0.220 & 0.213 - 0.217 & 0.215 - 0.218 & 0.211 - 0.217 & 1.61 - 1.63 & 1.58 - 1.60 & 1.59 - 1.61 & 1.52 - 1.56 \\ \midrule
    \multirow{5}{*}{0.49} & -2 & 35.0 - 35.2 & 34.9 - 35.1 & 34.9 - 35.1 & 35.0 - 35.2 & 0.213 - 0.216 & 0.215 - 0.218 & 0.211 - 0.215 & 0.211 - 0.215 & 1.57 - 1.59 & 1.58 - 1.60 & 1.56 - 1.58 & 1.55 - 1.58 \\
     & -4 & 35.1 - 35.3 & 35.0 - 35.1 & 35.0 - 35.1 & 35.0 - 35.2 & 0.213 - 0.218 & 0.215 - 0.219 & 0.213 - 0.217 & 0.212 - 0.215 & 1.58 - 1.61 & 1.60 - 1.63 & 1.57 - 1.60 & 1.57 - 1.60 \\
     & -6 & 35.1 - 35.2 & 34.9 - 35.1 & 35.0 - 35.1 & 35.0 - 35.2 & 0.214 - 0.218 & 0.215 - 0.218 & 0.212 - 0.216 & 0.212 - 0.215 & 1.58 - 1.61 & 1.60 - 1.62 & 1.57 - 1.60 & 1.57 - 1.59 \\
     & -8 & 35.1 - 35.1 & 35.0 - 35.1 & 34.9 - 35.1 & 35.0 - 35.1 & 0.215 - 0.216 & 0.216 - 0.219 & 0.212 - 0.216 & 0.212 - 0.217 & 1.59 - 1.61 & 1.60 - 1.62 & 1.57 - 1.60 & 1.57 - 1.61 \\
     & -10 & 35.0 - 35.2 & 35.0 - 35.2 & 35.0 - 35.1 & 35.0 - 35.1 & 0.215 - 0.218 & 0.214 - 0.219 & 0.212 - 0.215 & 0.211 - 0.215 & 1.59 - 1.61 & 1.58 - 1.61 & 1.57 - 1.59 & 1.56 - 1.59 \\ \midrule
    \multirow{5}{*}{0.62} & -4 & 35.9 - 36.2 & 35.9 - 36.1 & 35.9 - 36.1 & 36.4 - 36.6 & 0.211 - 0.215 & 0.212 - 0.215 & 0.212 - 0.216 & 0.211 - 0.217 & 1.57 - 1.60 & 1.57 - 1.59 & 1.57 - 1.60 & 1.59 - 1.63 \\
     & -6 & 35.4 - 35.5 & 35.3 - 35.4 & 35.4 - 35.5 & 36.5 - 36.6 & 0.212 - 0.216 & 0.212 - 0.215 & 0.213 - 0.216 & 0.211 - 0.215 & 1.57 - 1.60 & 1.57 - 1.60 & 1.58 - 1.60 & 1.61 - 1.64 \\
     & -8 & 35.5 - 35.6 & 35.3 - 35.5 & 35.3 - 35.6 & 36.6 - 36.8 & 0.213 - 0.218 & 0.212 - 0.216 & 0.214 - 0.217 & 0.212 - 0.216 & 1.58 - 1.62 & 1.56 - 1.59 & 1.58 - 1.60 & 1.62 - 1.65 \\
     & -10 & 35.6 - 35.8 & 35.4 - 35.5 & 35.4 - 35.6 & 36.7 - 36.8 & 0.215 - 0.218 & 0.212 - 0.216 & 0.214 - 0.217 & 0.212 - 0.215 & 1.59 - 1.61 & 1.57 - 1.59 & 1.57 - 1.59 & 1.62 - 1.65 \\
     & -12 & 35.6 - 35.8 & 35.4 - 35.6 & 35.5 - 35.6 & 36.8 - 36.9 & 0.216 - 0.219 & 0.212 - 0.216 & 0.213 - 0.216 & 0.212 - 0.217 & 1.59 - 1.61 & 1.56 - 1.59 & 1.57 - 1.59 & 1.62 - 1.65 \\ \bottomrule
    \end{tabular}
    \end{adjustbox}
\end{sidewaystable}

\newpage
\begin{sidewaystable}[h]
    \setlength{\tabcolsep}{6pt}
    \centering
    \caption{Range of the time-averaged values of $\beta$, $Fr_h$, and $Re_D$  (averaged at each $\lambda$) measured for each geometric configuration tested at $\beta = 45\%$.}
    \label{tab:flowVar45}
    \begin{adjustbox}{totalheight=0.5\textheight, center}
    \begin{tabular}{@{}|cc|cccc|cccc|cccc|@{}}
    \toprule
     &  & \multicolumn{4}{c|}{Measured $\beta$ [\%]} & \multicolumn{4}{c|}{Measured $Fr_h$} & \multicolumn{4}{c|}{Measured $Re_D$ [$\times10^-5$]} \\ \midrule
    $c/R$ & $\alpha_p$ [deg.] & $N=1$ & $N=2$ & $N=3$ & $N=4$ & $N=1$ & $N=2$ & $N=3$ & $N=4$ & $N=1$ & $N=2$ & $N=3$ & $N=4$ \\ \midrule
    \multirow{5}{*}{0.37} & 0 & 43.8 - 44.0 & 43.6 - 43.9 & 43.4 - 43.6 & 43.4 - 43.9 & 0.211 - 0.218 & 0.206 - 0.219 & 0.201 - 0.218 & 0.197 - 0.216 & 1.53 - 1.58 & 1.49 - 1.58 & 1.46 - 1.58 & 1.43 - 1.56 \\
     & -2 & 44.2 - 44.3 & 43.9 - 44.2 & 43.8 - 44.0 & 43.3 - 43.7 & 0.211 - 0.218 & 0.209 - 0.220 & 0.209 - 0.220 & 0.199 - 0.216 & 1.56 - 1.60 & 1.54 - 1.61 & 1.54 - 1.62 & 1.45 - 1.57 \\
     & -4 & 44.6 - 44.8 & 44.3 - 44.6 & 44.3 - 44.6 & 43.4 - 43.7 & 0.213 - 0.219 & 0.209 - 0.219 & 0.210 - 0.220 & 0.205 - 0.217 & 1.59 - 1.63 & 1.56 - 1.63 & 1.56 - 1.64 & 1.49 - 1.57 \\
     & -6 & 44.6 - 44.8 & 44.2 - 44.6 & 44.1 - 44.4 & 43.1 - 43.4 & 0.214 - 0.219 & 0.211 - 0.219 & 0.210 - 0.219 & 0.208 - 0.218 & 1.59 - 1.62 & 1.56 - 1.62 & 1.57 - 1.62 & 1.51 - 1.57 \\
     & -8 & 44.6 - 44.8 & 44.2 - 44.6 & 44.4 - 44.6 & 43.1 - 43.3 & 0.215 - 0.219 & 0.212 - 0.220 & 0.213 - 0.219 & 0.209 - 0.217 & 1.59 - 1.62 & 1.57 - 1.63 & 1.58 - 1.62 & 1.51 - 1.56 \\ \midrule
    \multirow{5}{*}{0.49} & -2 & 44.9 - 45.2 & 44.6 - 44.8 & 44.5 - 44.9 & 44.7 - 45.2 & 0.208 - 0.217 & 0.207 - 0.219 & 0.195 - 0.216 & 0.198 - 0.215 & 1.53 - 1.59 & 1.53 - 1.61 & 1.45 - 1.59 & 1.46 - 1.58 \\
     & -4 & 45.0 - 45.2 & 44.7 - 45.1 & 44.6 - 44.9 & 44.7 - 45.2 & 0.210 - 0.217 & 0.209 - 0.223 & 0.205 - 0.219 & 0.198 - 0.216 & 1.57 - 1.61 & 1.56 - 1.66 & 1.53 - 1.62 & 1.47 - 1.61 \\
     & -6 & 45.1 - 45.3 & 44.8 - 45.2 & 44.7 - 45.1 & 44.7 - 45.1 & 0.211 - 0.219 & 0.210 - 0.222 & 0.207 - 0.218 & 0.199 - 0.216 & 1.57 - 1.62 & 1.56 - 1.64 & 1.53 - 1.61 & 1.48 - 1.61 \\
     & -8 & 44.9 - 45.1 & 44.8 - 45.1 & 44.7 - 45.0 & 44.8 - 45.1 & 0.212 - 0.218 & 0.212 - 0.220 & 0.208 - 0.217 & 0.207 - 0.215 & 1.57 - 1.61 & 1.57 - 1.63 & 1.54 - 1.60 & 1.54 - 1.59 \\
     & -10 & 45.0 - 45.1 & 44.9 - 45.2 & 44.8 - 45.0 & 44.7 - 45.0 & 0.213 - 0.218 & 0.214 - 0.223 & 0.210 - 0.218 & 0.209 - 0.215 & 1.59 - 1.62 & 1.59 - 1.65 & 1.55 - 1.61 & 1.55 - 1.59 \\ \midrule
    \multirow{5}{*}{0.62} & -4 & 45.9 - 46.3 & 45.8 - 46.2 & 45.8 - 46.4 & 46.3 - 46.9 & 0.199 - 0.218 & 0.205 - 0.218 & 0.194 - 0.216 & 0.194 - 0.214 & 1.49 - 1.62 & 1.53 - 1.62 & 1.45 - 1.60 & 1.47 - 1.61 \\
     & -6 & 45.5 - 45.8 & 45.1 - 45.4 & 45.2 - 45.5 & 46.4 - 46.9 & 0.209 - 0.218 & 0.206 - 0.220 & 0.197 - 0.216 & 0.195 - 0.214 & 1.55 - 1.61 & 1.53 - 1.63 & 1.46 - 1.60 & 1.50 - 1.64 \\
     & -8 & 45.5 - 45.7 & 45.1 - 45.4 & 45.2 - 45.6 & 46.7 - 47.1 & 0.208 - 0.217 & 0.207 - 0.219 & 0.204 - 0.217 & 0.199 - 0.215 & 1.54 - 1.60 & 1.54 - 1.62 & 1.52 - 1.61 & 1.52 - 1.64 \\
     & -10 & 45.5 - 45.8 & 45.3 - 45.5 & 45.3 - 45.6 & 46.9 - 47.3 & 0.210 - 0.217 & 0.209 - 0.218 & 0.207 - 0.217 & 0.199 - 0.215 & 1.54 - 1.60 & 1.55 - 1.62 & 1.53 - 1.60 & 1.53 - 1.64 \\
     & -12 & 45.6 - 45.8 & 45.3 - 45.6 & 45.3 - 45.6 & 47.1 - 47.4 & 0.211 - 0.218 & 0.209 - 0.218 & 0.207 - 0.216 & 0.208 - 0.214 & 1.55 - 1.59 & 1.54 - 1.61 & 1.53 - 1.59 & 1.59 - 1.64 \\ \bottomrule
    \end{tabular}
    \end{adjustbox}
\end{sidewaystable}

\newpage
\begin{sidewaystable}[h]
    \setlength{\tabcolsep}{6pt}
    \centering
    \caption{Range of the time-averaged values of $\beta$, $Fr_h$, and $Re_D$ (averaged at each $\lambda$) measured for each geometric configuration tested at $\beta = 55\%$.}
    \label{tab:flowVar55}
    \begin{adjustbox}{totalheight=0.5\textheight, center}
    \begin{tabular}{@{}|cc|cccc|cccc|cccc|@{}}
    \toprule
     &  & \multicolumn{4}{c|}{Measured $\beta$ [\%]} & \multicolumn{4}{c|}{Measured $Fr_h$} & \multicolumn{4}{c|}{Measured $Re_D$ [$\times10^-5$]} \\ \midrule
    $c/R$ & $\alpha_p$ [deg.] & $N=1$ & $N=2$ & $N=3$ & $N=4$ & $N=1$ & $N=2$ & $N=3$ & $N=4$ & $N=1$ & $N=2$ & $N=3$ & $N=4$ \\ \midrule
    \multirow{5}{*}{0.37} & 0 & 53.4 - 53.7 & 53.3 - 53.7 & 53.1 - 53.7 & 52.9 - 53.6 & 0.214 - 0.220 & 0.212 - 0.221 & 0.211 - 0.221 & 0.207 - 0.219 & 1.54 - 1.58 & 1.54 - 1.60 & 1.53 - 1.60 & 1.51 - 1.59 \\
     & -2 & 54.1 - 54.5 & 54.5 - 55.0 & 53.6 - 54.0 & 52.8 - 53.4 & 0.216 - 0.222 & 0.215 - 0.225 & 0.211 - 0.221 & 0.207 - 0.219 & 1.58 - 1.62 & 1.57 - 1.63 & 1.55 - 1.62 & 1.51 - 1.60 \\
     & -4 & 54.4 - 54.6 & 55.1 - 55.6 & 54.1 - 54.7 & 52.8 - 53.5 & 0.214 - 0.219 & 0.216 - 0.226 & 0.210 - 0.222 & 0.208 - 0.219 & 1.60 - 1.64 & 1.59 - 1.66 & 1.56 - 1.64 & 1.51 - 1.59 \\
     & -6 & 54.3 - 54.6 & 54.0 - 54.7 & 54.0 - 54.6 & 52.5 - 53.0 & 0.216 - 0.220 & 0.212 - 0.221 & 0.211 - 0.221 & 0.208 - 0.218 & 1.61 - 1.64 & 1.58 - 1.64 & 1.57 - 1.64 & 1.51 - 1.58 \\
     & -8 & 54.3 - 54.6 & 54.1 - 54.6 & 54.0 - 54.6 & 52.5 - 53.1 & 0.216 - 0.220 & 0.212 - 0.220 & 0.212 - 0.222 & 0.208 - 0.220 & 1.60 - 1.62 & 1.57 - 1.63 & 1.57 - 1.64 & 1.50 - 1.58 \\ \midrule
    \multirow{5}{*}{0.49} & -2 & 54.8 - 55.2 & 54.6 - 55.1 & 54.5 - 55.2 & 54.4 - 55.1 & 0.213 - 0.220 & 0.213 - 0.222 & 0.208 - 0.219 & 0.204 - 0.218 & 1.57 - 1.62 & 1.57 - 1.63 & 1.53 - 1.61 & 1.51 - 1.60 \\
     & -4 & 54.9 - 55.5 & 54.7 - 55.3 & 54.5 - 55.2 & 54.6 - 55.4 & 0.212 - 0.220 & 0.213 - 0.222 & 0.207 - 0.220 & 0.205 - 0.219 & 1.57 - 1.63 & 1.58 - 1.65 & 1.55 - 1.64 & 1.53 - 1.62 \\
     & -6 & 55.1 - 55.5 & 54.7 - 55.2 & 54.6 - 55.3 & 54.6 - 55.3 & 0.214 - 0.219 & 0.212 - 0.222 & 0.209 - 0.221 & 0.206 - 0.218 & 1.58 - 1.62 & 1.58 - 1.65 & 1.55 - 1.64 & 1.53 - 1.61 \\
     & -8 & 55.2 - 55.5 & 54.6 - 55.1 & 54.5 - 55.0 & 54.7 - 55.1 & 0.216 - 0.221 & 0.212 - 0.222 & 0.209 - 0.219 & 0.207 - 0.218 & 1.59 - 1.62 & 1.57 - 1.64 & 1.55 - 1.62 & 1.53 - 1.62 \\
     & -10 & 55.0 - 55.5 & 54.8 - 55.2 & 54.6 - 55.2 & 54.5 - 55.1 & 0.216 - 0.221 & 0.213 - 0.222 & 0.210 - 0.220 & 0.207 - 0.218 & 1.59 - 1.62 & 1.57 - 1.64 & 1.55 - 1.62 & 1.53 - 1.61 \\ \midrule
    \multirow{5}{*}{0.62} & -4 & 56.5 - 56.8 & 56.0 - 56.8 & 55.8 - 56.6 & 56.5 - 57.4 & 0.211 - 0.219 & 0.207 - 0.222 & 0.205 - 0.221 & 0.202 - 0.217 & 1.57 - 1.63 & 1.54 - 1.64 & 1.53 - 1.64 & 1.53 - 1.63 \\
     & -6 & 55.6 - 55.9 & 55.2 - 55.9 & 55.9 - 56.7 & 56.7 - 57.5 & 0.212 - 0.220 & 0.209 - 0.221 & 0.208 - 0.224 & 0.203 - 0.216 & 1.57 - 1.63 & 1.55 - 1.63 & 1.54 - 1.64 & 1.56 - 1.66 \\
     & -8 & 55.1 - 56.0 & 55.3 - 55.9 & 56.0 - 56.8 & 57.1 - 57.8 & 0.212 - 0.219 & 0.209 - 0.222 & 0.208 - 0.223 & 0.209 - 0.218 & 1.57 - 1.62 & 1.55 - 1.63 & 1.53 - 1.63 & 1.57 - 1.66 \\
     & -10 & 55.7 - 56.1 & 55.4 - 56.1 & 56.0 - 56.5 & 57.2 - 58.0 & 0.212 - 0.219 & 0.210 - 0.223 & 0.210 - 0.223 & 0.206 - 0.219 & 1.56 - 1.62 & 1.56 - 1.65 & 1.54 - 1.63 & 1.58 - 1.67 \\
     & -12 & 55.7 - 56.2 & 55.3 - 55.9 & 55.2 - 56.1 & 57.5 - 58.1 & 0.213 - 0.220 & 0.211 - 0.220 & 0.208 - 0.220 & 0.207 - 0.219 & 1.57 - 1.62 & 1.56 - 1.63 & 1.54 - 1.62 & 1.58 - 1.67 \\ \bottomrule
    \end{tabular}
    \end{adjustbox}
\end{sidewaystable}

%% file: supp_uncertainty.tex
\section*{Uncertainty Analysis}

\begin{table}[hbt!]
    \centering
    \setlength{\tabcolsep}{6pt}
    \caption{Manufacturer-provided measurement uncertainty for the instruments used to measure turbine performance.}
    \label{tab:sensorInfo}
    \begin{tabular}{@{}ccc@{}}
        \toprule
        \textbf{Instrument} & \textbf{Measured Quantities} & \textbf{Measurement Uncertainty} \\ \midrule
        ATI Mini45-IP65 & $T$ [N], $L$ [N], $Q$ [N-m] & $\pm 1.25\%$ of full scale load \\ \midrule
        ATI Mini45-IP68 & $T$ [N], $L$ [N], $Q$ [N-m] & $\pm 1.25\%$ of full scale load \\ \midrule
        Motor Encoder & $\theta$ [deg.] & Resolution: $\pm 0.0055$ \\ \midrule
        Nortek Vectrino Profiler & $U_{\infty}$ [m/s] & $\pm 1\%$ of measured value $\pm 1$ mm/s \\ \midrule
        Omega LVU32 & $h$ [m] & Resolution: 0.25 mm
        \\ \bottomrule
        Omega Ultra-Precise RTD & Temp [$^{\circ}$C] & Display resolution: 0.1 $^{\circ}$C
    \end{tabular}
\end{table}

In previous work in this test facility, \citet{snortland_cycletocycle_2023} and \citet{hunt_experimental_2024} applied the American Society of Mechanical Engineers' standard on test uncertainty \citep{american_society_of_mechanical_engineers_test_2005} to similar test set ups in the same test facility.
In both cases, the authors found that the observed variability in turbine performance is significantly narrower than the uncertainty predicted based on formal analysis.
This is due to manufacturer-specified measurement uncertainty of the ATI load cells (\Cref{tab:sensorInfo}), which are conservative relative to their performance in practice.
Because of this, the variability in turbine performance is actually dominated by fluctuations in the freestream flow.

\begin{figure}[hbt!]
    \centering
    \begin{minipage}[t]{0.45\textwidth}
            \centering
            \includegraphics[width=\textwidth]{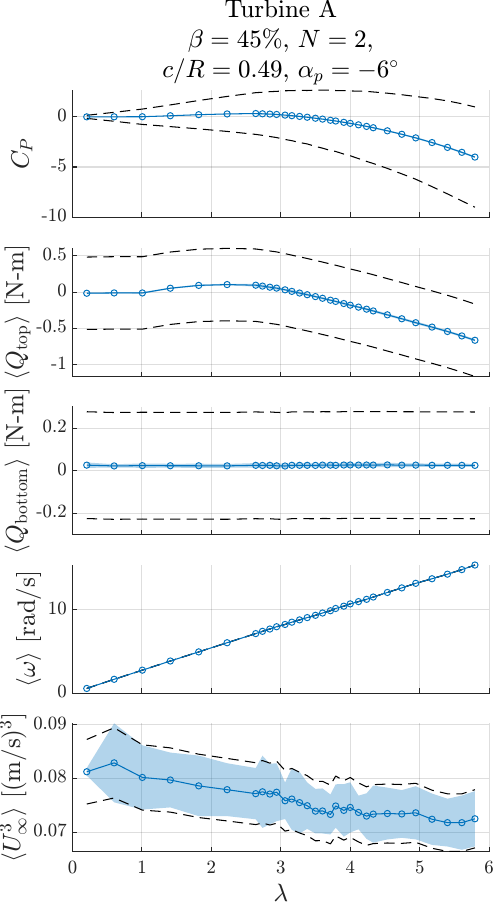}
    \end{minipage} \hfill
    \begin{minipage}[t]{0.45\textwidth}
            \centering
            \includegraphics[width=\textwidth]{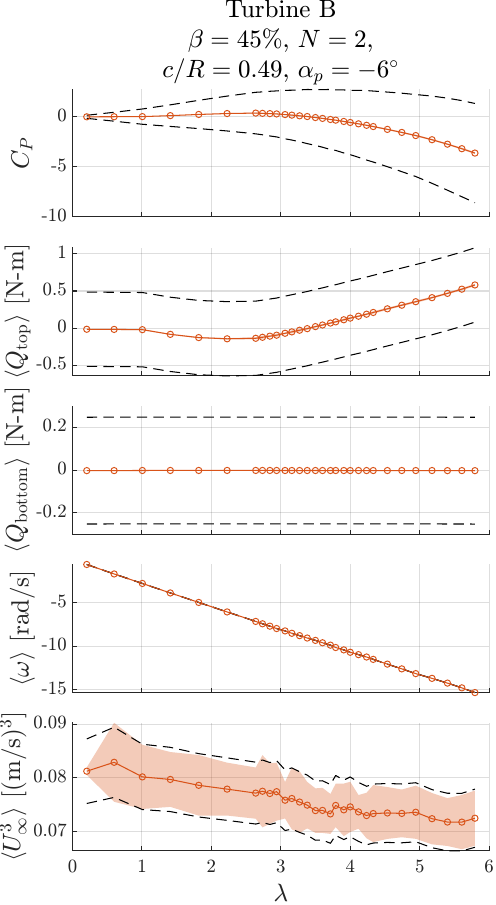}.
    \end{minipage}
    \caption{Uncertainty of $\langle C_P \rangle$ and component measurements for both turbines in a representative experiment. The dashed lines indicate the 95\% confidence intervals obtained via the standard for test uncertainty, whereas the shaded blue region indicates $\pm 2$ standard deviations of the measured cycle-average values. For all but $\langle U_{\infty}^3 \rangle$, the shaded region is on the order of the line width. For $\langle \omega \rangle$, the 95\% confidence interval is also on the order of the line width.}
    \label{fig:uncertainty}
\end{figure}

Uncertainty analysis on the experimental data from the current study yields an identical result, as shown for a representative case in \cref{fig:uncertainty}.
This is expected, since the flume setup and individual turbine test rigs in this study are nearly identical to that used by \citet{hunt_experimental_2024} (a different ADV and load cell models are used here, but from the same manufacturers).
Consequently, we refer the reader to prior work \citep{snortland_cycletocycle_2023,hunt_experimental_2024} for further discussion on the experimental uncertainty of this test set-up.

%% file: supp_lateral.tex
\section*{Lateral Force Coefficients}

\begin{figure*}[hbt!]
    \centering
    \includegraphics[width = \textwidth]{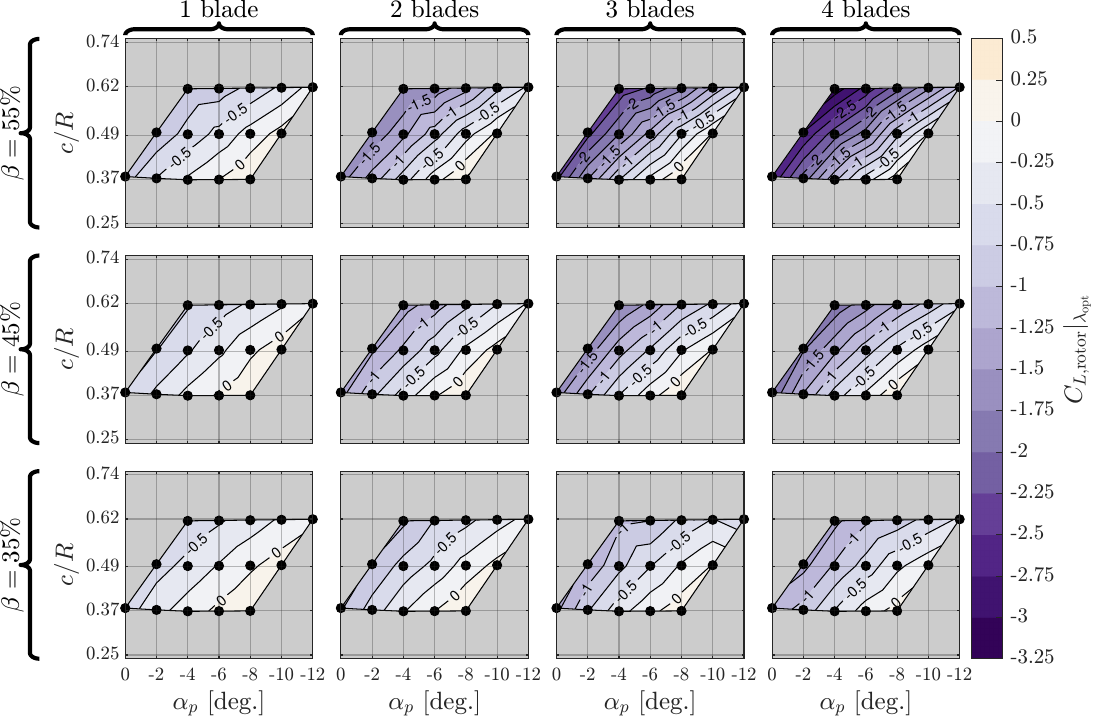}
    \caption{Contours of the time-average rotor-average lateral coefficient at maximum array performance as a function of $c/R$ and $\alpha_p$ at each combination of $\beta$ and $N$ tested.}
    \label{fig:clHeatmap_array}
\end{figure*}

As noted in \Cref{methods:arrayLayout}, since this cross-flow turbine array is symmetric about the channel centerline and operating with $\Delta \theta = 0^{\circ}$, the lateral forces experienced by the turbines in the array are equal and opposite.
Consequently, while the array-average thrust coefficient is approximately 0 at all $\lambda$ across all experiments, the lateral forces experienced by the individual turbines are non-zero and may change with the blockage ratio and rotor geometry.
Therefore, we define the directions of $L_{A}$ and $L_{B}$ in \cref{fig:arrayOverhead} to account for this counter-rotation, such that the array-averaged lateral force coefficient instead represents the average lateral force coefficient for an individual turbine in the array (i.e., $C_{L,\mathrm{rotor}}$).

\Cref{fig:clHeatmap_array} shows the contours of the time-averaged rotor-average lateral force coefficient at maximum efficiency (i.e., $C_{L,\mathrm{rotor}}\rvert_{\lambda_{\mathrm{opt}}}$) across the tested parameter space.
Trends in $C_{L,\mathrm{rotor}}\rvert_{\lambda_{\mathrm{opt}}}$ with blockage are similar to those of $C_{P,\mathrm{array}}$ and $C_{T,\mathrm{array}}\rvert_{\lambda_{\mathrm{opt}}}$, in that $C_{L,\mathrm{rotor}}\rvert_{\lambda_{\mathrm{opt}}}$ increases with blockage for all geometries, and increases more rapidly for geometries with larger solidity.
Other trends in $C_{L,\mathrm{rotor}}\rvert_{\lambda_{\mathrm{opt}}}$ match those observed at for a single turbine at $\approx11\%$ blockage by \citep{hunt_experimental_2024}.
$C_{L,\mathrm{rotor}}\rvert_{\lambda_{\mathrm{opt}}}$ exhibits a strong dependence on the preset pitch angle, and for all tested $\beta\!-\!N$ combinations, $C_{L,\mathrm{rotor}}\rvert_{\lambda_{\mathrm{opt}}}$ increases with increasing $c/R$ and less-negative $\alpha_p$ (i.e., more toe-in preset pitch).
Lower lateral force with more negative preset pitch is consistent with trends in the array-average thrust coefficients with $\alpha_p$ (\Cref{fig:ctHeatmap}), where more toe-out preset pitch reduced the forcing on the array.

%% file: supp_cylinders.tex
\section*{Cylinder Thrust Coefficients}

\begin{figure*}[hbt!]
    \centering
    \includegraphics[width = 0.6\textwidth]{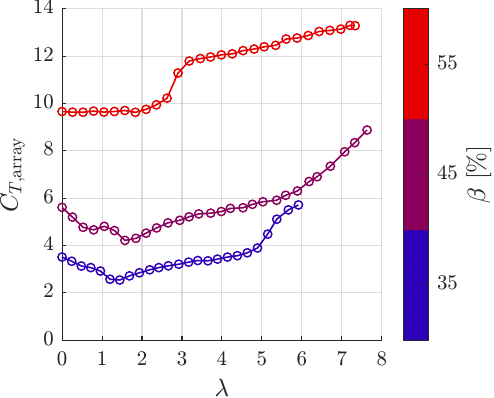}
    \caption{Time-averaged array-average thrust coefficient versus tip-speed ratio for the cylindrical shells as a function of $\beta$.}
    \label{fig:ctCylinder}
\end{figure*}

%% file: supp_scaling.tex
\section*{Implications of Bluff-Body Scaling for Individual Geometries}

\begin{figure*}[hbt!]
    \centering
    \includegraphics[width = \textwidth]{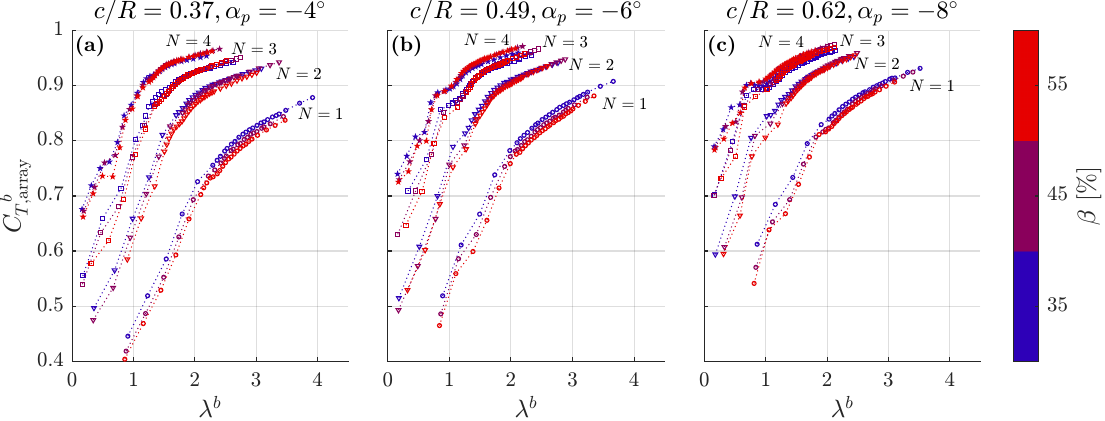}
    \caption{$C_{T,\mathrm{array}}^{\ \: b}$ vs $\lambda^b$ at various $\beta$ and $N$ for rotor geometries with \textbf{(a)} $c/R = 0.37$ and $\alpha_p = -4^{\circ}$, \textbf{(b)} $c/R = 0.49$ and $\alpha_p = -6^{\circ}$, and \textbf{(c)} $c/R = 0.49$ and $\alpha_p = -8^{\circ}$.}
    \label{fig:ctTSRBluff}
\end{figure*}

While \Cref{fig:bluffBodyScaling} highlights high-level trends in $C_{T,\mathrm{array}}^{\ \: b}$ with ${\sigma_d}^b$ across the tested parameter space, how these trends inform self-similarity in array thrust across $\beta$ for specific geometric configurations is considered in \Cref{fig:ctTSRBluff}, which shows the bypass-scaled characteristic thrust curves (i.e, $C_{T,\mathrm{array}}^{\ \: b}$ vs $\lambda^b$) for several representative geometries.
Higher solidity geometries tend to exhibit better agreement in the $C_{T,\mathrm{array}}^{\ \: b}-\lambda^b$ curves across $\beta$ (e.g., $N=1$ vs $N=2$ for $c/R = 0.49$ and $\alpha_p = -6^{\circ}$ in \Cref{fig:ctTSRBluff}b), and this collapse begins at lower values of $\lambda^b$ for higher solidity geometries.
Furthermore, at sufficiently high solidity, $C_{T,\mathrm{array}}^{\ \: b}$ appears to become less sensitive to $\sigma$ (e.g., $N=3$ and $N=4$ for $c/R = 0.62$ and $\alpha_p = -8^{\circ}$ in \Cref{fig:ctTSRBluff}c).
Taken together, these results imply that, as dynamic solidity increases, \citeauthor{maskell_ec_theory_1963}-inspired bluff-body scaling is expected to become more effective at capturing self-similarity in array performance across blockage ratios, which is supported by the aggregate trends in \Cref{fig:bluffBodyScaling}b.
However, we note that at higher ${\sigma_d}^b$, $C_{T,\mathrm{array}}^{\ \: b}$ at $\beta = 35\%$ begins to diverge from that of $\beta = 45\%$ and $55\%$, although the magnitude of this difference is relatively small.